\documentclass[%
 reprint,
 floatfix,
 amsmath,amssymb,
 aps,
prb,
]{revtex4-2}

\usepackage{comment}
\usepackage{graphicx}
\usepackage{dcolumn}
\usepackage{bm}
\usepackage[colorlinks=true,linkcolor=blue,citecolor=blue,urlcolor=blue]{hyperref}

\usepackage[usenames,dvipsnames]{xcolor}

\begin{document}

\title{Signatures of triplet correlations in density of states of Ising superconductors}

\author{M. Haim}
\affiliation{The Racah Institute of Physics, The Hebrew University of Jerusalem, Jerusalem 9190401, Israel.}
\author{D. M\"ockli}
\affiliation{The Racah Institute of Physics, The Hebrew University of Jerusalem, Jerusalem 9190401, Israel.}
\author{M. Khodas}
\affiliation{The Racah Institute of Physics, The Hebrew University of Jerusalem, Jerusalem 9190401, Israel.}

\begin{abstract}
The few-layer transition metal dichalcogenides (TMDs) have been recently suggested as a platform for controlled  unconventional superconductivity. 
We study the manifestations of unconventional triplet pairing in the density of states of a disordered TMD based monolayer. 
The conventional singlet pairing attraction is assumed to be dominant paring interaction.
We map the phase diagrams of disordered Ising superconductors in the plane of temperature and the in-plane magnetic field. 
The latter suppresses singlet and promote  triplet correlations. 
The triplet order parameters of a trivial (non-trivial) symmetry compete (cooperate) with singlet order parameter which gives rise to a rich phase diagram.
We locate the model dependent phase boundaries and compute the order parameters in each of the distinct phases.  
With this information, we obtain the density of states by solving the Gorkov equation.
The triplet components of the order parameters may change an apparent width of the density of states by significantly increasing the critical field.
The triplet components of the order parameters lead to the density of states broadening significantly exceeding the broadening induced by magnetic field and disorder in singlet superconductor.

\end{abstract}

\maketitle

\section{\label{sec:Intro}Introduction}

The progress in growth and fabrication techniques made it possible to create high quality ultra-thin multi-layer systems with individual layers held together by weak Van der Waals forces \cite{Geim2013}.
In particular, few- and monolayers of transition metal dichalcogenides (TMD)s with highly tunable properties have been fabricated \cite{Wang2012}.
Many such systems turned out to be superconducting down to the monolayer limit \cite{Lu2015,Ugeda2015,Saito2016,Xi2016,Costanzo2016,Dvir2017,DelaBarrera2018,Sohn2018} greatly stimulating the research of two-dimensional superconductivity.
Very recently, the TMD based systems have been proposed as a platform for controlled studies of the intrinsic or externally induced unconventional superconductivity \cite{Hamill2020,Cho2020}.

Many TMD monolayers such as 1H-NbSe$_2$ lack the inversion center even though the bulk (2H-NbSe$_2$) has such a center.
This gives rise to a spin splitting of electron bands in the presence of atomic spin-orbit interaction. 
Due to the basal mirror plane symmetry $\sigma_h$, the electron spins are polarized out-of-plane.
The superconducting properties of such systems referred to as Ising superconductors \cite{Yuan2014,Lu2015,Saito2016,Xi2016} are determined to a large extent by the spin splitting, $\Delta_{\mathrm{SO}}$ typically exceeding the superconducting gap by few orders of magnitude.

One of the  experimentally confirmed signatures of Ising superconductivity  is its  remarkable stability to the in-plane magnetic field, $\mathbf{B} \perp \hat{z}$. 
The in-plain critical field, $B_c$ is demonstrated to greatly exceed the Pauli limit \cite{Lu2015,Saito2016,Xi2016,Dvir2017,DelaBarrera2018,Sohn2018,Liu2018}. 
In fact, $B_c$ is infinite at zero temperature, $T=0$ unless the disorder is present in the system \cite{Bulaevskii976,Sosenko2017,Ilic2017,Mockli2020} or 
a random Rashba spin-orbit coupling is produced by ripples breaking $\sigma_h$ symmetry \cite{HuertasHernando2006}.
Hereinafter, we absorb a half product of a $g$-factor and the Bohr magneton in the definition of $B$ such that the Zeeman spin splitting is $2 B$.
In few-layer systems we neglect the coupling of the in-plane field to the orbital motion.
Such coupling is dominant for out-of plane field  \cite{Mineev2007}.

A large $B_c$ can be explained by an in-plane spin susceptibility being close to that of a normal state \cite{Xi2016}.
The superconductivity is then stabilized as the magnetic polarization energy is excluded from the energy difference between the normal and the superconducting states.
In contrast to the case of conventional superconductors, the electrons with momenta in between the spin split Fermi surfaces reorient their spins in response to an applied in-plane field \cite{Wickramaratne2020}.
The net spin polarization along the field is independent of $\Delta_{\mathrm{SO}}$ and is determined by the Pauli susceptibility in the normal state.

A smooth adjustment of paired electrons to the applied field is secured by a $\sigma_h \mathcal{T}$ symmetry 
combining $\sigma_h$ with the time reversal symmetry $\mathcal{T}$ \cite{Fischer2018}. 
At finite $\mathbf{B}$ the  spins of paired electrons related to each other by $\sigma_h \mathcal{T}$ operation are no longer anti-parallel.
It follows that as electrons are polarized by the field, the wave function of the Cooper pairs they form inevitably acquires a triplet component \cite{Mockli2018}.
Such field induced triplets, here referred to as non-trivial, have a symmetry lower than the symmetry of the crystal.
Therefore, it is meaningful to assign the two distinct transition temperatures, $T_{\mathrm{cs}}$ and $T_{\mathrm{ct}}$ to the leading singlet and subleading non-trivial triplet interaction channels, respectively.

Crucially, the non-trivial triplets are distinct by symmetry from the triplets coexisting with singlets in the absence of inversion center at $\mathbf{B}=0$ \cite{Gorkov2001,Yip2014a,Smidman2017}.
We refer to the latter triplets as trivial as they transform trivially under all symmetry operations.
In the limit studied here $\Delta_{\mathrm{SO}}\ll E_F$, where $E_F$ is the Fermi energy, trivial triplets decouple from singlets \cite{Frigeri2004d}.
For this reason, we characterize a possible attraction in a Cooper pairs forming a trivial triplet by a separate transition temperature, $T_{\mathrm{ctz}}$. 
The trivial triplets are insensitive to moderate fields and are suppressed by a minute disorder \cite{Mockli2020}.
In contrast, non-trivial triplets are induced by the field and are stabilized against the disorder due to the strong coupling to the leading singlet order parameter (OP) \cite{Mockli2019,Mockli2020}.

For the trivial triplets to be observed the critical temperature should be large enough, $T_{\mathrm{ctz}} \lesssim T_{\mathrm{cs}}$ as otherwise the singlet correlations dominate at experimentally accessible fields. 
In contrast, the non-trivial triplets noticeably affect the $B_c$ and OPs already for $T_{\mathrm{ct}} \ll T_{\mathrm{cs}}$.
In both cases, for triplet correlations to come into play, electrons forming a triplet Cooper pair should attract.

The indirect evidence for attraction in triplet channel comes from the very recent Density Functional Theory calculations performed on the NbSe$_2$ monolayers either free standing or on a substrate.
These studies find a large Stoner enhancement of the magnetic susceptibility indicative of strong ferromagnetic interactions \cite{Wickramaratne2020} and/or ferromagnetic ground state \cite{Divilov2020}.

The ferromagnetic fluctuations revealed by Density Functional Theory enhance the pairing interaction in the triplet channel and suppress the interactions in the singlet channel \cite{Fay1977,Sigrist2005,Samokhin2008,Mineev2017}. 
As argued in Ref.~\cite{Wickramaratne2020} significant attraction in the triplet channel results from ferromagnetic fluctuations with correlation length exceeding $v_F/\Delta_{\mathrm{SO}}$, where $v_F$ is the Fermi velocity and we set $\hbar= k_{\mathrm{B}}=1$. 
In the alternative scenario considered in Ref.~\cite{Shaffer2020} appropriate to the gated TMDs with small Fermi pockets both singlet and triplet instabilities arise from repulsion. In this approach $T_{\mathrm{cs}}$ and $T_{\mathrm{ct}} = T_{\mathrm{ctz}}$ are promoted by distinct inter-pocket pair hopping processes, and can be both finite.

\begin{figure*}%[htp]
\centering
\includegraphics[width=0.96\textwidth]{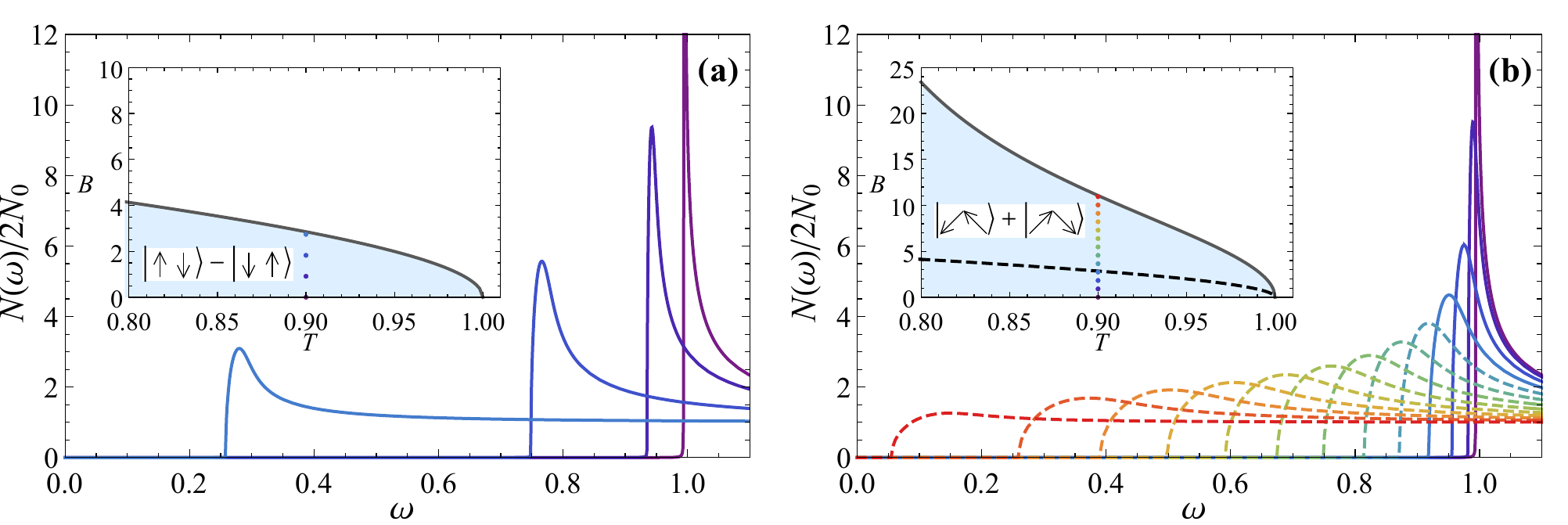}
\caption{\label{fig:res1} 
The DOS, $N\left(\omega\right)$ normalized by its normal state value, $2N_{0}$ as a function of $\omega$ given in units of $T_{\mathrm{cs}}$.
In this example the SOC is given by $K$-model (Sec.~\ref{sec:model}). 
$\Delta_{\mathrm{SO}}=15T_{\mathrm{cs}}$, $\Gamma=0.1T_{\mathrm{cs}}$. 
Panel (a): singlet OP, $T_{\mathrm{ct}}=0$. 
Panel (b): singlet-triplet OP, $T_{\mathrm{ct}}=0.8 T_{\mathrm{cs}}$.
Insets: $(T,B)$ phase diagrams with $T$ and $B$ in units of $T_{\mathrm{cs}}$.
The solid (grey) line shows, $B_c(T)$ separating the normal and superconducting phase shown as shaded (blue) area.
The coordinates of a vertically aligned and evenly spaced dots, $(T=0.9,B)$ are the $(T,B)$ pairs for which the DOS curves are shown in the main figure using the same color scheme.
The dashed black line in the inset of panel (b) is the $B_c(T)$ shown as a solid line in panel (a) for singlet only case. 
In panels (a) and (b) DOS curves broaden with increasing field.
}
\end{figure*}

Here we assume finite $T_{\mathrm{ct(z)}} < T_{\mathrm{cs}}$ and study the effect of triplet correlations on the density of states (DOS), $N(\omega)$, where $\omega$ is the quasi-particle energy.
Typically, the DOS curve, $N(\omega)$ is inferred from the low-temperature differential conductance in devices contacted via a tunnel barrier \cite{Dvir2017}. 
A recent tunneling data in gated MoS$_2$ is in fact, indicative of an unconventional symmetry of the OP in gated TMDs \cite{Costanzo2018}.

The essential conclusion of this work is that an admixture of a triplet component to the superconducting OP may result in apparent broadening of the $N(\omega)$ curves as the applied field increases. 
Indeed, as detailed in Sec.~\ref{sec:qualit} for $\Delta_{\mathrm{SO}} \ll E_F$,  the scalar disorder characterized by the scattering rate $\Gamma$ acts as a spin flipping disorder with the effective rate,
$\Gamma_{\mathrm{eff}} \propto \Gamma B^2/\Delta^2_{\mathrm{SO}}$ in the experimentally relevant regime, $B \lesssim \Delta_{\mathrm{SO}}$. 
This holds also in the opposite limit, $\Delta_{\mathrm{SO}} \gtrsim E_F$, Ref.~\cite{Sosenko2017}.
The triplet components describing the Cooper pairs with parallel spins do not add to broadening per se.
In other words, when the triplet component of OP is added and all the other  parameters are kept fixed the DOS broadening inferred from the Gorkov equation does not change appreciably. 
Nevertheless, triplet components push the critical field $B_c(T)$ up.
This, in turn, leads to a stronger broadening $\propto B_c^2$ even in systems that are nominally in the clean limit, $\Gamma \ll T_{\mathrm{cs}}$, see Fig.~\ref{fig:res1}.

We may express this point yet differently (see Sec.~\ref{sec:Discussion}).
At fixed $B_c(T)$, addition of the triplet components of OP enforces the adjustment of the other model parameters via self-consistency condition. And this leads to extra broadening compared with the pure singlet OP.

These results are illustrated in Fig.~\ref{fig:res1}, where the computed DOS is compared for systems with purely singlet OP (panel (a)) achieved for $T_{\mathrm{ct(z)}}=0$ and the OP which contains a field induced triplet correlations present for $T_{\mathrm{ct}} > 0$.
Qualitatively, as the Fig.~\ref{fig:res1} demonstrates, a broadening growing rapidly towards $B_c$ in very clean systems with $\Gamma = 0.1 T_{\mathrm{cs}}$ might be an indication of a finite triplet component of the OP.

The field dependence of the DOS broadening is a salient feature of Ising superconductors, along with the enhanced $B_c$.
As such it applies equally  to pure singlet and to the mixed parity superconductors. 
Its concrete manifestation, however, differs in these two cases.
To highlight these differences which are of a potential experimental importance 
we investigate in detail the $(T,B)$ mean field phase diagrams of a monolayer TMD for a subset of a relevant symmetry constrained OPs. 
The phase diagrams are constructed for the complimentary models of the nodal and nodeless  $\Delta_{\mathrm{SO}}$, and for systems with different degree of purity.
In each case the representative DOS curve is presented.

The paper is structured as follows. 
In Sec. \ref{sec:model} we specify the model Hamiltonian including the kinetic energy, interaction constrained by symmetries and the disorder potential.
Section \ref{sec:Results} contains a summary of main results intended for the reader not interested in the details of the derivation.
In this section, we show phase diagrams for selected sets of parameters as well as few representative DOS curves computed along different lines on these phase diagrams.
The Gorkov equation employed for finding the DOS is detailed in Sec.~\ref{sec:GE}. 
Qualitative picture of interrelation between the triplet correlations and the DOS is given in Sec.~\ref{sec:qualit}.
The Landau expansion of the thermodynamic potential we rely upon in order to map the phase diagram is obtained in Sec.~\ref{sec:order_parameters}. 
Finally in Sec.~\ref{sec:Discussion} we discuss the results in light of the recent tunneling experiments.

\section{The model Hamiltonian}
\label{sec:model}

Consider a disordered monolayer superconductor without an inversion center.
The appropriate Hamiltonian 
\begin{align}\label{H}
H = H_0 + H_{dis} + H_{int}    
\end{align}
includes the kinetic energy, random disorder potential and pairing interaction term, respectively.  
Kinetic energy,
\begin{align}\label{H0}
H_0 =\underset{\mathbf{k},s}{\sum}\xi_{\mathbf{k}}c_{\mathbf{k}s}^{\dagger}c_{\mathbf{k}s}
+\underset{\mathbf{k},ss'}{\sum}\left[\boldsymbol{\gamma}\left(\mathbf{k}\right)-\mathbf{B}\right]\!\cdot\boldsymbol{\sigma}_{ss'}c_{\mathbf{k}s}^{\dagger}c_{\mathbf{k}s'}
\end{align}
contains the dispersion measured from the chemical potential, $\xi\left(\mathbf{k}\right)$,
the anti-symmetric spin-orbit  coupling (SOC), $\boldsymbol{\gamma}\left(-\mathbf{k}\right)=-\boldsymbol{\gamma}\left(\mathbf{k}\right)$ due to the lack of the inversion center, and Zeeman field $\mathbf{B}=B\hat{x}$, Fig.~\ref{fig:crystal}.
We denote, 
$c_{\mathbf{k}s}^{\dagger}=V^{-1/2}\int d\mathbf{r}e^{i\mathbf{k}\cdot\mathbf{r}}\psi_{\mathbf{r}s}^{\dagger}$, 
where $\psi_{\mathbf{r}s}^{\dagger}$ creates a particle with spin projection $s$ on the $z$-axis at position $\mathbf{r}$ in a volume $V$.
The vector of Pauli matrices is denoted by $\boldsymbol{\sigma}=\left(\sigma_{1},\sigma_{2},\sigma_{3}\right)$, and $\sigma_0$ stands for a unit matrix in spin space.

In this work we consider the two models of the band structure and SOC both having $D_{3h}$ as the point symmetry group, Fig.~\ref{fig:crystal}a.
NbSe$_2$ monolayer the Nb derived band crossing the Fermi level gives rise to the two distinct hole Fermi pockets.
In the hexagonal Brillouin Zone one of the pockets is centered at $\Gamma$ and the other two disconnected pockets enclose the $\pm K$ points, Fig.~\ref{fig:crystal}b.
Although both pockets are present in NbSe$_2$, here for simplicity we consider two separate models referred to as $\Gamma$- and $K$-model with only one type of pockets.  

The $\sigma_h$ symmetric SOC has a form $\boldsymbol{\gamma}\left(\mathbf{k}\right) = \Delta_{\mathrm{SO}}\hat{\gamma}\left(\mathbf{k}\right)\hat{z}$.
As the axial vector $\hat{z}$ belongs to $A'_2$ irrep of $D_{3h}$, the same should be true for the scalar function $\hat{\gamma}\left(\mathbf{k}\right)$.
For simplicity we write $\hat{\gamma}\left(\mathbf{k}\right) = \hat{\gamma}\left(\varphi_\mathbf{k}\right)$, where $\varphi_\mathbf{k}$ is an angle the vector $\mathbf{k}$ forms with the $k_x$ axis in the Brillouin Zone.
The acceptable functions $\hat{\gamma}\left(\varphi_\mathbf{k}\right)$ are linear combinations of Fourier harmonics $\cos(3 n \varphi_\mathbf{k})$ with integer $n \neq 0$.
All such functions vanish along $\Gamma M$ as prescribed by the vertical mirror symmetry. 
Hence, without loss of generality we take for the $\Gamma$-model 
$\hat{\gamma}\left(\varphi_\mathbf{k}\right) = \sqrt{2} \cos( 3 \varphi_\mathbf{k} )$ \cite{Sigrist1991,Smidman2017}.
The simplest model of SOC for $K$-model is $\hat{\gamma}\left(\varphi_\mathbf{k}\right) = \mathrm{sgn}[\cos( 3 \varphi_\mathbf{k}) ]$.
Such a function is constant at each of the $\pm K$ pockets.
We considered normalized $\hat{\gamma}$ functions, 
$\left\langle  \hat{\gamma}^2\left(\varphi_\mathbf{k}\right)  \right\rangle_{\mathrm{F}} =1 $, where the angular averaging over the Fermi surface is denoted as 
$\langle \cdots \rangle_{\mathrm{F}}\equiv (2\pi)^{-1}\int d\varphi_{\mathbf{k}}(\cdots )$.

The disorder is modeled as a collection of impurities of a density $n_{imp}$ located at random positions, $\mathbf{R}_j$. 
The resulting scattering potential reads, 
\begin{align}
\label{eq:Hdis}
H_{dis} = \sum_j \sum_{s=1,2} \sum_{\mathbf{k},\mathbf{k}'} u_{\mathbf{k}-\mathbf{k}'} e^{i \mathbf{R}_j \cdot (\mathbf{k}-\mathbf{k}')}    c_{\mathbf{k}s}^{\dagger}c_{\mathbf{k}'s}\, ,
\end{align}
where $u_{\mathbf{q}}$ is a Fourier transformation of the spin independent, scalar potential produced by a single impurity.
For simplicity we consider a short range impurity potential such that $u_{\mathbf{q}} = u_0$.
The resulting elastic scattering time, $\tau$ is given by the Golden Rule, 
$\tau^{-1} = 2 \pi n_{imp} N_0 u_0^2$, 
where $N_0$ is the normal state DOS per spin species.

Finally, we treat the pairing interaction
\begin{align}\label{Hint}
H_{int}  =\frac{1}{2}\underset{\mathbf{k},\mathbf{k}'}{\sum}\underset{\left\{ s_{i}\right\} }{\sum}V_{s_{1}'s_{2}'}^{s_{1}s_{2}}\left(\mathbf{k},\mathbf{k}'\right)
c_{\mathbf{k}s_{1}}^{\dagger}c_{-\mathbf{k}s_{2}}^{\dagger}c_{-\mathbf{k}'s_{2}'}c_{\mathbf{k}'s_{1}'}
\end{align}
within the mean field approximation.
To this end we introduce the OP 
\begin{equation}
\Delta_{s_{1}s_{2}}\left(\mathbf{k}\right)=\frac{1}{V}\underset{\mathbf{k}',s_{1}',s_{2}'}{\sum}V_{s_{1}'s_{2}'}^{s_{1}s_{2}}\left(\mathbf{k},\mathbf{k}'\right)
\left\langle c_{-\mathbf{k}',s_{2}'}c_{\mathbf{k}',s_{1}'}\right\rangle 
\label{eq:self}
\end{equation}
and make an approximation, 
$H_{\mathrm{int}} \approx H_{\mathrm{MF}} - \bar{H}_{i}$, where
the mean field interaction Hamiltonian is
\begin{align}
\label{eq:MF}
H_{\mathrm{MF}} & =\frac{V}{2}\underset{\mathbf{k},s_{i}}{\sum}
\Delta_{s_{1}s_{2}}^{*}\left(\mathbf{k}\right)c_{-\mathbf{k}s_{2}}c_{\mathbf{k}s_{1}} +h.c.\, ,
\end{align}
where $h.c.$ stands for Hermitian conjugated term.
Equation \eqref{eq:self} is a self-consistency equation with the right hand side computed with the quadratic Hamiltonian \eqref{eq:MF}.
The expectation value of the mean field Hamiltonian,
\begin{align}
\bar{H}_{i} = \frac{1}{2}\underset{s_{i},s_{i}'}{\sum}\underset{\mathbf{k},\mathbf{k}'}{\sum}V_{s_{1}'s_{2}'}^{s_{1}s_{2}}\left(\mathbf{k},\mathbf{k}'\right)\left\langle c_{\mathbf{k},s_{1}}^{\dagger}c_{-\mathbf{k},s_{2}}^{\dagger}\right\rangle\left\langle c_{-\mathbf{k}',s_{2}'}c_{\mathbf{k}',s_{1}'}\right\rangle. 
\end{align}

Next, we introduce a simplified model of interaction, \eqref{Hint} which contains a minimum amount of necessary information to capture the thermodynamic properties and field induced phase transitions in the superconducting TMD monolayer.
To this end we invoke the arguments based on symmetry considerations.

\subsection{Symmetry and the choice of the OPs}
We now consider the momentum and spin dependence of the superconducting OP written in the standard form as 
\begin{equation}
\label{eq:OP}
\Delta\left(\mathbf{k}\right)=\left[\psi\left(\mathbf{k}\right)\sigma_{0}+\mathbf{d}\left(\mathbf{k}\right)\cdot\boldsymbol{\sigma}\right]i\sigma_{2}.
\end{equation}
Here $\psi\left(\mathbf{k}\right)$ and $\mathbf{d}\left(\mathbf{k}\right)$ parametrize the singlet and triplet components of the OP. 

In our description the leading OP is singlet, $T_{\mathrm{cs}} > T_{\mathrm{ct(z)}}$. 
Here we neglect it's anisotropy setting $\psi\left(\mathbf{k}\right)=\eta_0 \hat{\psi}_{0}$ with the basis function $\hat{\psi}_{0}=1$. 
At $\mathbf{B} =0$ the triplet OP coexisting with the singlet one is determined by the axial vector
$\mathbf{d}_{A'_1}\left(\mathbf{k}\right)$ transforming as $A_1'$.
Since $\mathbf{d}\left(\mathbf{k}\right)$ and the SOC transform in the same way, 
the reasoning fixing $\boldsymbol{\gamma}(\mathbf{k})$ applies, and we write  
$\mathbf{d}_{A_1'}\left(\mathbf{k}\right) = \eta_{A} \hat{\gamma}\left(\mathbf{k}\right)\hat{z}$.
In addition to $\mathbf{d}_{A_1'}$ we include the triplet OP induced by the field, previously introduced in Ref.~\cite{Mockli2019}.
Since the in-plane field $\mathbf{B} =(B_x, B_y)$ belongs to $E''$ it couples to the triplet OPs of the same symmetry.
For the $\Gamma$-model within the considered space of the Fourier harmonics the pairs of $E''$-partners
$(\mathbf{d}_{E''}^{1},\mathbf{d}_{E''}^{2})$ are $(  \cos 3 \varphi_\mathbf{k}\hat{x},   \cos 3 \varphi_\mathbf{k}\hat{y} )$, $( \cos \varphi_\mathbf{k}\hat{x} -  \sin \varphi_\mathbf{k}\hat{y} ,  \sin \varphi_\mathbf{k} \hat{x} +  \cos \varphi_\mathbf{k}\hat{y} )$, \cite{Hamill2020} and $(  \sin 3 \varphi_\mathbf{k}\hat{x},  \sin 3 \varphi_\mathbf{k} \hat{y})$.

The first OP in the full list above written in the form $( \hat{\gamma}(\varphi_\mathbf{k}) \hat{x} , \hat{\gamma}(\varphi_\mathbf{k})\hat{y})$ applies to both $\Gamma$- and $K$-models.
In contrast to other $E''$ combinations, it couples to the fields via the SOC induced polarization of the bands even when the interaction has a full rotational symmetry such that different Fourier harmonics decouple.
Although in the generic situation all $E''$ triplets condense together, here we consider the model with rotational invariant interaction. 
This narrows the list of $E''$ triplet OPs down to one entry,   
$(\mathbf{d}_{E''}^{1},\mathbf{d}_{E''}^{2}) = \left[\eta_{E1}  \hat{\gamma}(\varphi_\mathbf{k})\hat{x} , \eta_{E2}\hat{\gamma}(\varphi_\mathbf{k})\hat{y} \right]$.
We do not include the triplet OPs given by the $\mathbf{d}$-vectors
$  \cos \varphi_\mathbf{k}\hat{x} +  \sin \varphi_\mathbf{k}\hat{y}$,
$  \sin 3  \varphi_\mathbf{k}\hat{z} $, 
$\sin \varphi_\mathbf{k}\hat{x}  -  \cos \varphi_\mathbf{k}\hat{y}$, and
belonging to $A''_{1}$, $A'_2$ and $A''_2$ symmetry, respectively.
We also omit the $E'$ triplet assuming no strain.  

\begin{figure}[htp]
\centering
\includegraphics[width=0.48\textwidth]{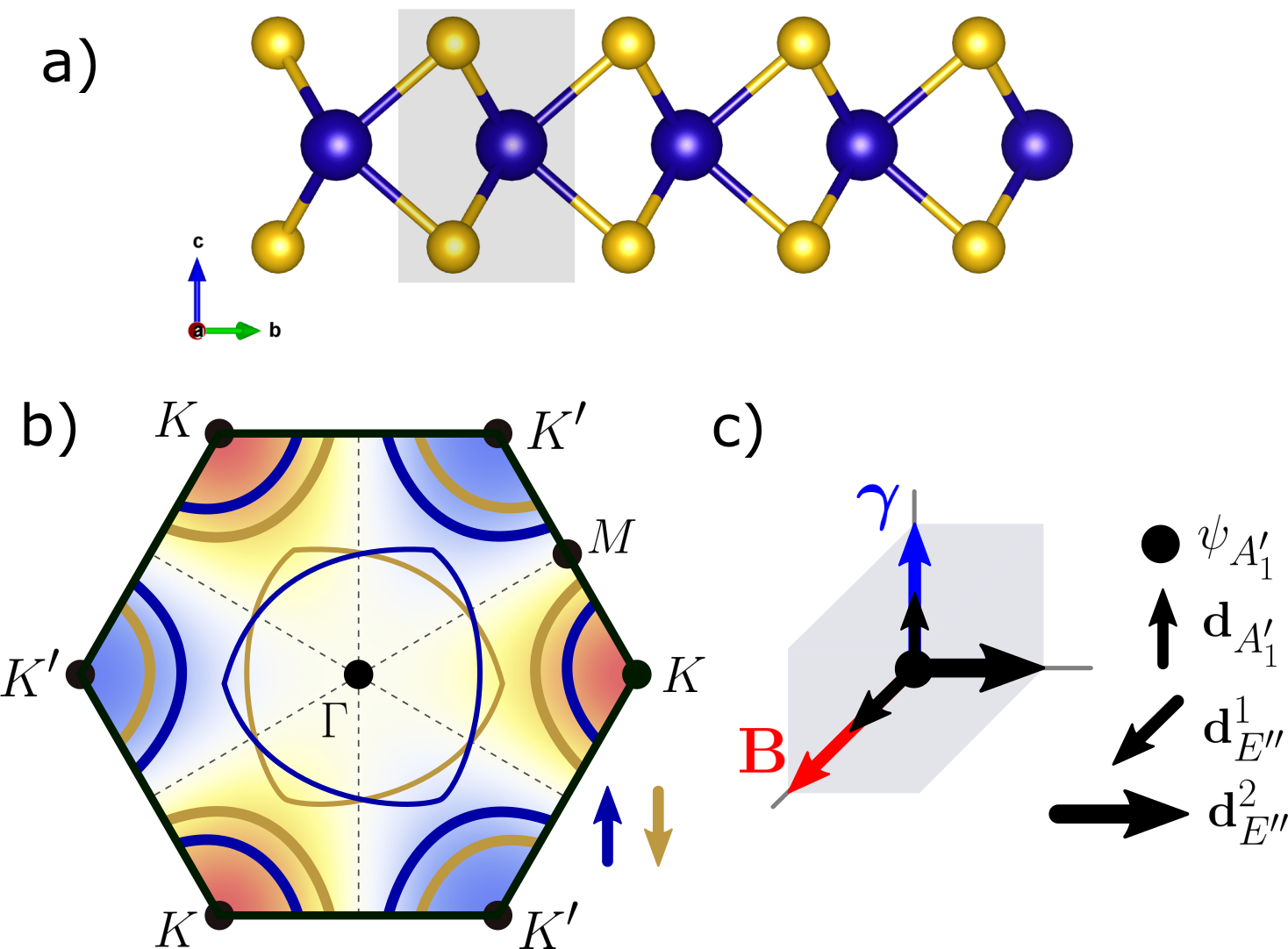}
\caption{\label{fig:crystal}
a) The crystal structure of 1H-NbSe$_2$, where the Nb are blue and Se yellow. The gray region shows the unit cell in which there is no inversion center. This figure was made with Vesta \cite{Momma2011}. b) Schematic Fermi surface of metallic monolayer TMDs at zero magnetic field. SOC vanishes along the dashed $\Gamma M$ lines. The sign of SOC alternates between adjacent dashed lines. c) The directions of the Zeeman field (red), SOC (blue) and the OPs (black).}
\end{figure}

In summary, in our model the $\mathbf{d}$-vector characterizing the triplet component of the OP, Eq.~\eqref{eq:OP} reads,
\begin{equation}
\label{eq:d}\mathbf{d}\left(\mathbf{k}\right)=\hat{\gamma}\left(\mathbf{k}\right)
\left(  \eta_{E1}\hat{x} + \eta_{E2} \hat{y} + \eta_{A}\hat{z}\right)\, ,  
\end{equation}
and our list of OPs includes $A'_1$ singlet $\eta_0$,
$A'_1$ triplet $\eta_A$, and $E''$ triplet $(\eta_{E1},\eta_{E2})$, see Fig~\ref{fig:crystal}c.

The above symmetry arguments lead us to the effective interaction 
\begin{align}\label{eq:Hint_eff}
V_{s_{1}'s_{2}'}^{s_{1}s_{2}}\left(\mathbf{k},\mathbf{k}'\right)&=  v_{s}\left[i\sigma_{2}\right]_{s_{1}s_{2}}\left[i\sigma_{2}\right]_{s_{1}'s_{2}'}^{*}
\\
+&\underset{j=1,2}{\sum}v_{t}\left[\hat{\gamma}\left(\mathbf{k}\right)\sigma_{j}i\sigma_{2}\right]_{s_{1}s_{2}}
\left[\hat{\gamma}\left(\mathbf{k}'\right)\sigma_{j}i\sigma_{2}\right]_{s_{1}'s_{2}'}^{*}\nonumber\\
+& v_{tz}\left[\hat{\gamma}\left(\mathbf{k}\right)\sigma_{3}i\sigma_{2}\right]_{s_{1}s_{2}}\left[\hat{\gamma}\left(\mathbf{k}'\right)\sigma_{3}i\sigma_{2}\right]_{s_{1}'s_{2}'}^{*}.
\nonumber 
\end{align}
Equation \eqref{eq:Hint_eff} is the minimal Hamiltonian capturing the interaction in the $A'_{1}$ singlet channel of a strength, $v_s$, the interaction in the $A'_{1}$ triplet channel of a strength $v_{tz}$, and finally the $E''$ channel with interaction $v_{t}$.

With the model Hamiltonian, \eqref{eq:Hint_eff} the OP, Eq.~\eqref{eq:OP} 
is fully determined by the four OPs of a definite symmetry,
\begin{align}\label{eq:MF_symm}
\eta_0 & = \frac{v_s}{V} 
\underset{\mathbf{k}',s_{1}'s_{2}'}{\sum} 
\left[i\sigma_{2}\right]_{s_{1}'s_{2}'}^{*} 
\left\langle c_{-\mathbf{k}',s_{2}'}c_{\mathbf{k}',s_{1}'}\right\rangle
\\
\eta_{E1(2)} & = \frac{v_t}{V}
\underset{\mathbf{k}',s_{1}'s_{2}'}{\sum}
\left[\hat{\gamma}\left(\mathbf{k}'\right)\sigma_{1(2)}i\sigma_{2}\right]_{s_{1}'s_{2}'}^{*}
\left\langle c_{-\mathbf{k}',s_{2}'}c_{\mathbf{k}',s_{1}'}\right\rangle
\notag \\
\eta_{A} & = \frac{v_{tz}}{V}
\underset{\mathbf{k}',s_{1}'s_{2}'}{\sum}
\left[\hat{\gamma}\left(\mathbf{k}'\right)\sigma_{3}i\sigma_{2}\right]_{s_{1}'s_{2}'}^{*}
\left\langle c_{-\mathbf{k}',s_{2}'}c_{\mathbf{k}',s_{1}'}\right\rangle \notag \, .
\end{align}

It is convenient to use the observable transition temperatures than the interaction amplitude.
We hence introduce the three transition  temperatures $T_{\mathrm{cs}}$, $T_{\mathrm{ct(z)}}$ corresponding to the three terms in Eq.~\eqref{eq:Hint_eff}.
These temperatures are defined under conditions that only one interaction amplitude is non-zero, the system is clean, SOC and magnetic field are turned off.
In this case, we have the standard relations, 
$T_{\mathrm{cs}} = 2  \Lambda e^{\gamma_E}\pi^{-1} \exp( - 1/2 N_0 |v_s| )$, and
$T_{\mathrm{ct(z)}} = 2  \Lambda e^{\gamma_E}\pi^{-1} \exp( - 1/2 N_0 |v_{t(z)}| )$ where $\Lambda$ is a high energy cutoff for the attraction and $\gamma_E$ is Euler's constant \cite{Kita2015}.

\begin{figure*}[htp]
\centering
\includegraphics[width=1\textwidth]{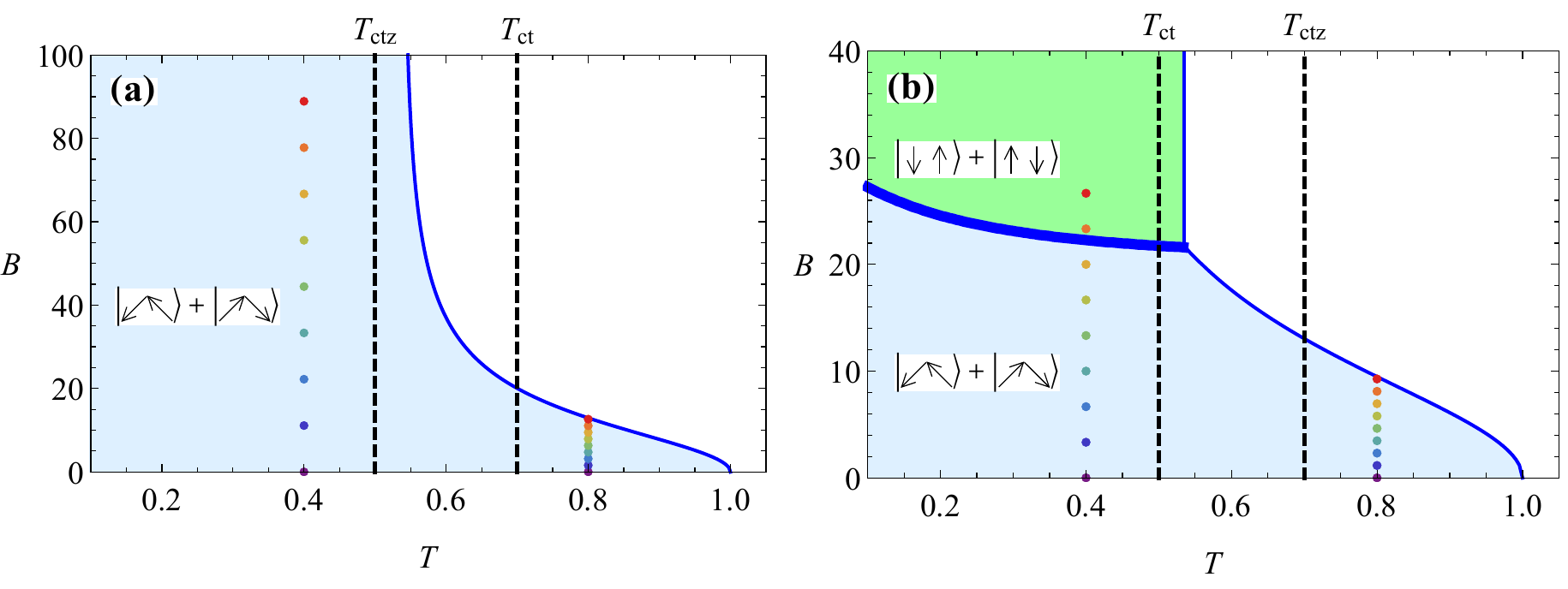}
\caption{\label{fig:phase_d1} The $(T,B)$ phase diagrams with both axes given in units of $T_{\mathrm{cs}}$ for the $K$-model with $ \Delta_{\mathrm{SO}}=15 T_{\mathrm{cs}},\Gamma=0.2T_{\mathrm{cs}}$.
The light shaded (blue) regions mark the superconducting phase of coexisting singlet and $\eta_{E2}$ triplet OP ($s+if$ phase). 
The second order normal to superconducting transition lines are denoted by thin (blue) lines.
The $T_{\mathrm{ct}}$ and $T_{\mathrm{ctz}}$ are marked by vertical dashed lines.
(a)  $T_{\mathrm{ct}}=0.7 T_{\mathrm{cs}}$, $T_{\mathrm{ctz}}=0.5 T_{\mathrm{cs}}$. 
(b) $T_{\mathrm{ct}}=0.5 T_{\mathrm{cs}}$, $T_{\mathrm{ctz}}=0.7 T_{\mathrm{cs}}$. 
The shaded (green) region in panel (b) marks superconducting phase with $\eta_{A}$ triplet OP ($s'$ phase). 
The first order transition between $s+if$ and $s'$ phases is denoted by a thick (blue) line.
The coordinates of vertically aligned and evenly spaced colored dots, at $T= 0.4T_{\mathrm{cs}} $ and $T= 0.8T_{\mathrm{cs}}$ in panel (a)[(b)] define the $(T,B)$ pairs for which the DOS is shown
in Fig.~\ref{fig:DOS1}(a,b) [Fig.~\ref{fig:DOS1}(c,d)], respectively,
using the same colors scheme.  
}
\end{figure*}

\section{Summary of results}
\label{sec:Results}

\begin{figure*}[htp]
\centering
\includegraphics[width=1\textwidth]{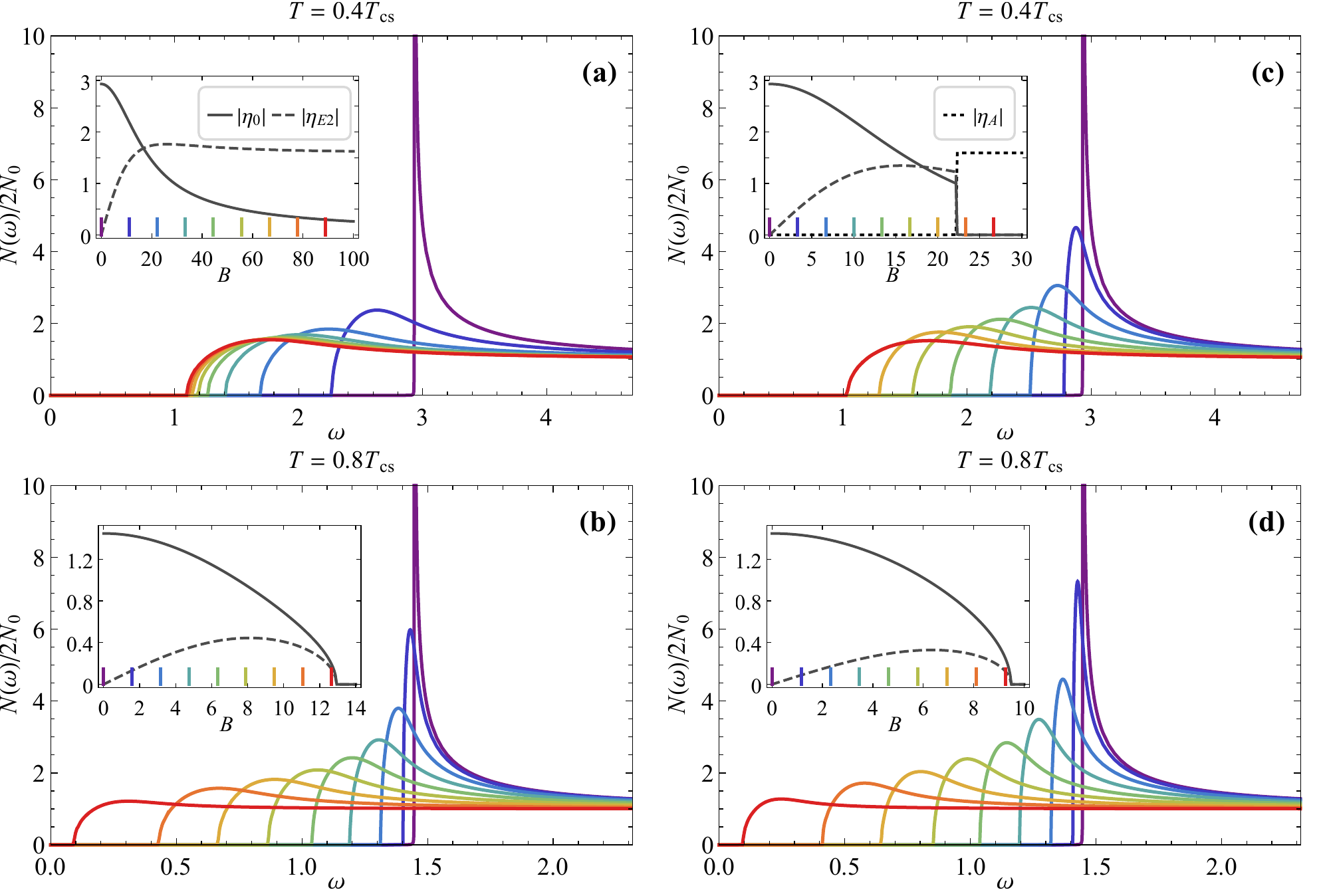}
\caption{\label{fig:DOS1} Normalized DOS, $N\left(\omega\right)/2N_{0}$. 
Panels (a),(b) pertain to $T=0.4T_{\mathrm{cs}}$ and $T=0.8T_{\mathrm{cs}}$ columns of dots in Fig.\ref{fig:phase_d1}(a), respectively. 
Panels (c),(d) pertain to $T=0.4T_{\mathrm{cs}}$ and $T=0.8T_{\mathrm{cs}}$ columns of dots in Fig.\ref{fig:phase_d1}(b), respectively. 
The DOS curves have the same color as the corresponding dots in Fig.\ref{fig:phase_d1}, and broaden with increasing field. 
Insets: the evolution of OPs with $B$. 
The magnitude of singlet $\eta_0$, triplet $\eta_{E2}$ and triplet $\eta_{A}$ OPs are shown by solid, dashed and dotted lines, respectively.
The values of $B$ for which the DOS is plotted in the main figure is marked by a vertical bar of the same color as the DOS curve attached to a horizontal axis.
OPs, $\omega$ and $B$ are in units of $T_{\mathrm{cs}}$.}
\end{figure*} 

\begin{figure*}[htp]
\centering
\includegraphics[width=1\textwidth]{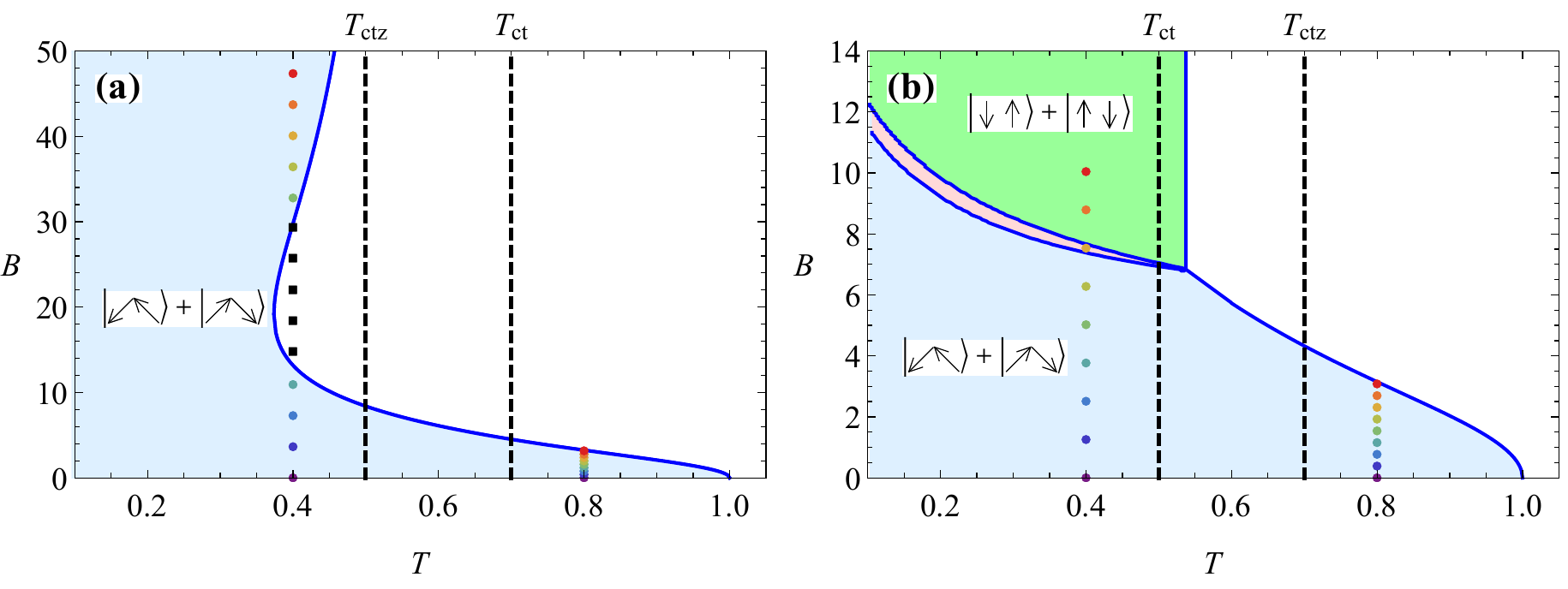}
\caption{\label{fig:phase_d2} The $(T,B)$ phase diagrams with both axes given in units of $T_{\mathrm{cs}}$ for the $\Gamma$-model with $ \Delta_{\mathrm{SO}}=15 T_{\mathrm{cs}},\Gamma=0.2T_{\mathrm{cs}}$.
The light shaded (blue) regions mark the superconducting phase of coexisting singlet and $\eta_{E2}$ triplet OP ($s+if$ phase). 
The second order normal to superconducting transition lines are denoted by thin (blue) lines.
The $T_{\mathrm{ct}}$ and $T_{\mathrm{ctz}}$ are marked by vertical dashed lines.
(a)  $T_{\mathrm{ct}}=0.7 T_{\mathrm{cs}}$, $T_{\mathrm{ctz}}=0.5 T_{\mathrm{cs}}$. 
(b) $T_{\mathrm{ct}}=0.5 T_{\mathrm{cs}}$, $T_{\mathrm{ctz}}=0.7 T_{\mathrm{cs}}$. 
The shaded (green) region in panel (b) marks superconducting phase with $\eta_{A}$ triplet OP ($s'$ phase).
The thin elongated (red) region in between the $s+if$ and $s'$ phases denotes an intermediate $s+i f +is'$ phase.
The coordinates of vertically aligned and evenly spaced colored dots, at $T= 0.4T_{\mathrm{cs}} $ and $T= 0.8T_{\mathrm{cs}}$ in panel (a)[(b)] define the $(T,B)$ pairs for which the DOS is shown 
in Fig.~\ref{fig:DOS2}(a,b) [Fig.~\ref{fig:DOS2}(c,d)], respectively, using the same colors scheme.  
}
\end{figure*}

\begin{figure}[htp]
\includegraphics[scale=0.65]{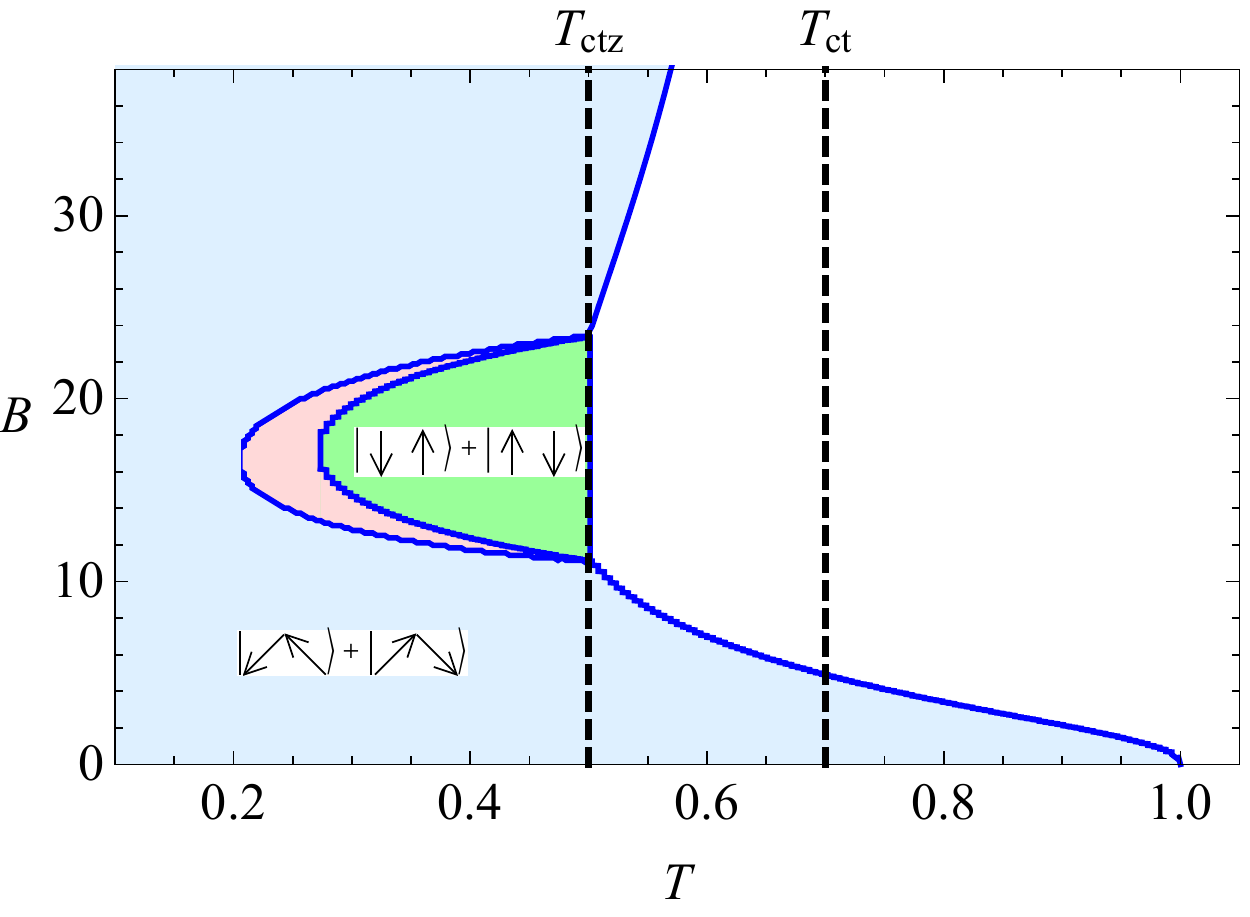}
\caption{\label{fig:phase_d3} 
The $(T,B)$ phase diagrams with both axes given in units of $T_{\mathrm{cs}}$ for the $\Gamma$ model with $\Delta_{\mathrm{SO}}=15 T_{\mathrm{cs}}$ and no disorder.
The $T_{\mathrm{ctz}}=0.5 T_{\mathrm{cs}}$, $T_{\mathrm{ct}}=0.7 T_{\mathrm{cs}}$,
are indicated by vertical dashed lines.
The light shaded (blue) regions mark the superconducting phase of coexisting singlet and $\eta_{E2}$ triplet OP ($s+if$ phase). 
The shaded (green) region marks superconducting phase with $\eta_{A}$ triplet OP ($s'$ phase). 
The thin crescent shaped (red) region in between the $s+if$ and $s'$ phases denotes an intermediate $s+i f +is'$ phase. 
The normal state occupies the high-$T$ part of the phase diagram. 
All the phase boundaries shown as solid (blue) lines mark the second order phase transitions. 
}
\end{figure}

\begin{figure*}[htp]
\centering
\includegraphics[width=1\textwidth]{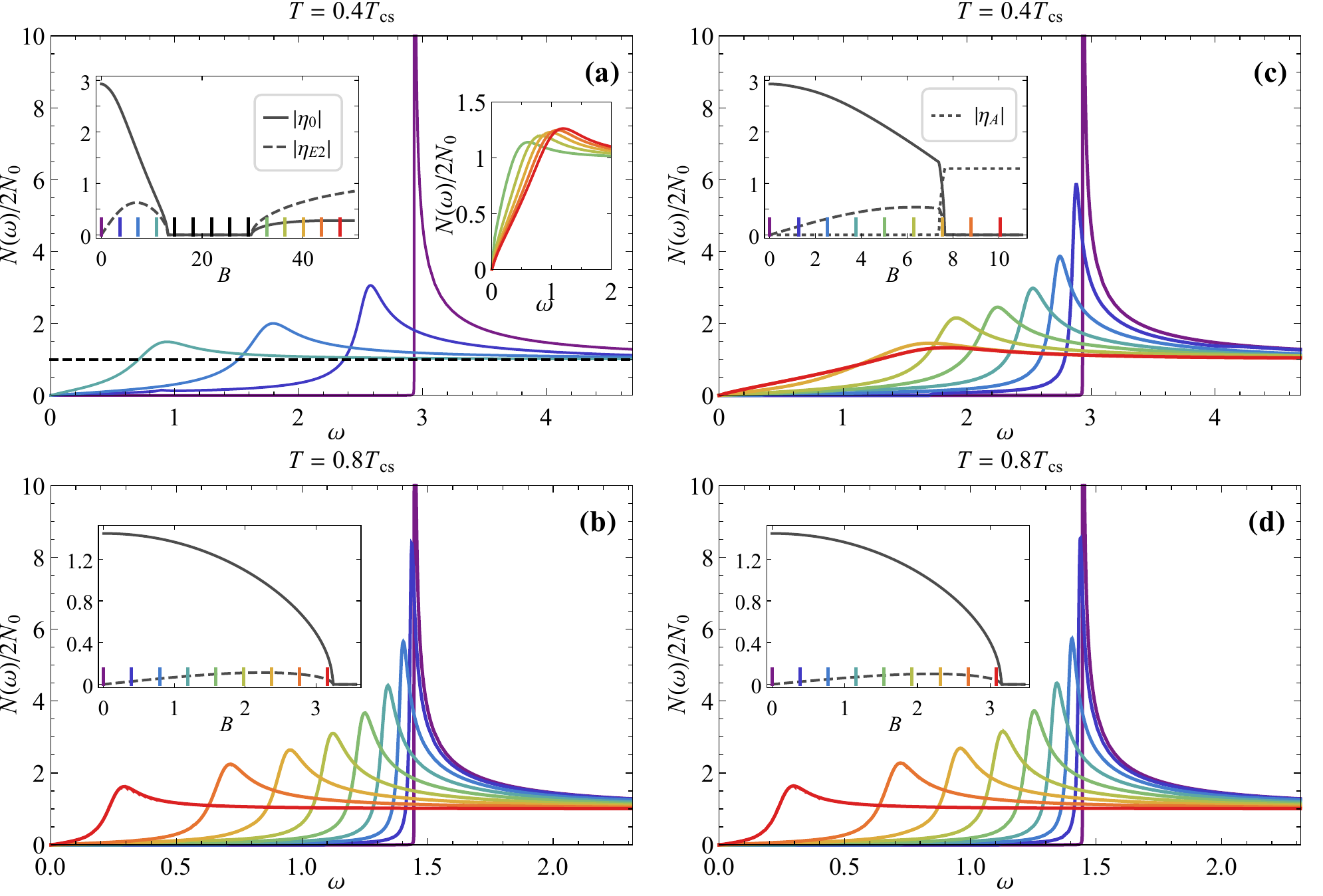}
\caption{\label{fig:DOS2} Normalized DOS, $N\left(\omega\right)/2N_{0}$ for the $\Gamma$-model. 
Panels (a),(b) pertain to $T=0.4T_{\mathrm{cs}}$ and $T=0.8T_{\mathrm{cs}}$ columns of dots in Fig.\ref{fig:phase_d2}(a), respectively. 
Panels (c),(d) pertain to $T=0.4T_{\mathrm{cs}}$ and $T=0.8T_{\mathrm{cs}}$ columns of dots in Fig.\ref{fig:phase_d2}(b), respectively. 
%%%%
The DOS curves have the same color as the corresponding dots in Fig.\ref{fig:phase_d2}, and broaden with increasing field. 
The horizontal (black) dashed line in panel (a) is the DOS for normal state points in the phase diagram in Fig.\ref{fig:phase_d2}(a) indicated by black squares. 
Panel (a): The right inset shows the high-field DOS. 
All panels: the left inset shows the evolution of OPs with field.
The magnitude of singlet $\eta_0$, triplet $\eta_{E2}$ and triplet $\eta_{A}$ OPs are shown by solid, dashed and dotted lines, respectively.
OPs, $\omega$ and $B$ are in units of $T_{\mathrm{cs}}$.
Panel (b): The critical field in the Pauli limit is $0.86T_{\mathrm{cs}}$ and the gape closes at $B^*\approx1.4T_{\mathrm{cs}}$.}
\end{figure*}

In this section we present the main findings of this work.
The end result is the calculated DOS of a disordered Ising superconductor throughout the $(T,B)$ phase diagram. 
More specifically, we focused on the role played by the triplet correlations.
Although the $\Gamma$- and $K$-model of SOC give overall similar results, there are qualitative differences in some regions of the phase diagram.
We, therefore, consider these two models separately.

Furthermore, for each of the two models the two physically distinct situations arise depending on the symmetry of the dominant triplet channel.
The first scenario for which the field induced $E''$ triplet channel dominates over the symmetric $A_1'$ triplet channel is realized for $T_{\mathrm{ct}} > T_{\mathrm{ctz}}$.
And the opposite scenario is realized for $T_{\mathrm{ctz}} > T_{\mathrm{ct}}$.
Below we consider these two scenarios separately. 
\subsection{The \texorpdfstring{$K$}{}-model of SOC}
\subsubsection{\texorpdfstring{$T_{\mathrm{ct}} > T_{\mathrm{ctz}}$}{}: dominant \texorpdfstring{$E''$}{} triplet channel}
\label{sec:ct>ctz}
The representative phase diagram in the situation when $E''$ triplet channel dominate the $A_1'$ triplets, is shown in Fig.~\ref{fig:phase_d1}a.
In this case there is a normal and superconducting phases with the superconducting OP having both $A_1'$ singlet and $E''$ field-induced triplet components.
In the limit, $\Delta_{\mathrm{SO}}\ll E_F$, considered here, the $A_1'$ triplet component remains zero.
Previously, we have referred to this phase as having the $s+if$ symmetry \cite{Mockli2019}. 
The critical field diverges at a finite temperature $T_{\infty}$.
This temperature can be easily computed by noticing that in the high field limit, the singlet component of the OP is suppressed.
The triplet component of OP, on the other hand, has a $E''$ symmetry with $\mathbf{d} \parallel \hat{y}$ which is perpendicular to $\mathbf{B}$, see Fig.~\ref{fig:crystal}c.
Therefore, magnetic field does not limit this triplet OP.
Although $\mathbf{d}$ in this case also orthogonal to $\boldsymbol{\gamma}(\mathbf{k})$ in the high field limit,
$B \gg \Delta_{\mathrm{SO}}$ the limiting of the $E''$ triplet by SOC is eliminated.
As a result, in the considered limit $T_{\infty}$ is determined by the standard relation,
\begin{align}
\label{eq:Tinf}
\ln\frac{T_{\infty}}{T_{\mathrm{ct}}} = \Psi\left(\frac{1}{2}\right) - \Psi\left(\frac{1}{2}+\frac{\Gamma}{2 \pi T_{\infty}}\right)  \, ,
\end{align}
where $\Psi$ is the digamma function.
Clearly, $T_{\infty}=T_{\mathrm{ct}}$ in the clean system and is suppressed by the disorder.
For instance, for the set of parameters used in Fig.~\ref{fig:phase_d1}a Eq.~\eqref{eq:Tinf} gives $T_{\infty} \approx 0.536$.
At $\Gamma > \pi T_{\mathrm{ct}}e^{-\gamma}/2$, $B_c$ is finite for all temperatures, and the high field part of the superconducting phase is eliminated.
Much stronger disorder is needed to bring $B_c$ to the Pauli limit, see Ref.~\cite{Mockli2020} for details. 

Both OPs and DOS are plotted along two constant temperature cuts in the phase diagram Fig.~\ref{fig:DOS1}a and Fig.~\ref{fig:DOS1}b, respectively.
Low $T$ cut, Fig.~\ref{fig:DOS1}a shows the saturation of the triplet OP and vanishing of singlet OP at high field confirming the statements made above.
The DOS similarly is broadened with increasing the field and saturates in the same limit.
At higher $T$ the DOS shows a strong smearing effect of the field, while the gap is still visible, Fig.~\ref{fig:DOS1}b.
\subsubsection{\texorpdfstring{$T_{\mathrm{ctz}} > T_{\mathrm{ct}}$}{}: dominant \texorpdfstring{$A_{1}'$}{} triplet channel}

Fig.~\ref{fig:phase_d1}b features a phase diagram of the system with $A_{1}'$ triplet pairing dominating $E''$ triplet correlations.
In this case the low field second order transition line bifurcates at the tricritical point as the field increases.
Among the two high field transition lines emanating from the tricritical point one marks the first-order phase transition between the $s+if$ state and the $s'$ state where the OP is a pure $A_1'$ symmetric triplet.
The other is a vertical line, $T=T_v$  of the second order transitions between the normal state and $s'$ superconducting state.
The transition temperature $T_v$ is determined by the instability towards the $A_1'$ triplet.
For this triplet the $\mathbf{d} \parallel \hat{z}$, see Fig.~\ref{fig:crystal}c which is parallel to $\mathbf{\gamma}(\mathbf{k})$ and orthogonal to $\mathbf{B}$.
This implies that the $A'_{1}$ transition temperature, $T_v$ is insensitive neither to SOC nor to the magnetic field, and is given by a standard equation,
\begin{align}
\label{eq:Tv}
\ln\frac{T_v}{T_{\mathrm{ctz}}} = \Psi\left(\frac{1}{2}\right) - \Psi\left(\frac{1}{2}+\frac{\Gamma}{2 \pi T_v}\right)  \, ,
\end{align}
where $\Psi$ is the digamma function.
For the parameters used in Fig.~\ref{fig:phase_d1}a, Eq.~\eqref{eq:Tv} gives $T_v \approx 0.536$.
Note that $T_v$ accidentally coincides with $T_{\infty}$ found in Sec.~\ref{sec:ct>ctz}.
More generally, the temperature at which the critical field diverges is determined by  $\max\{T_{\mathrm{ct}},T_{\mathrm{ctz}}\}$.
For both Fig.~\ref{fig:phase_d1}a and Fig.~\ref{fig:phase_d1}b $\max\{T_{\mathrm{ct}},T_{\mathrm{ctz}}\} =0.7 T_{\mathrm{cs}}$.
For this reason for the current set of critical temperatures, $T_v = T_{\infty}$.
Another manifestation of the similarity of the two triplet instabilities at high fields 
can be seen by comparison of the OP field dependence  in Fig.~\ref{fig:DOS1}a and Fig.~\ref{fig:DOS1}c at $T=0.4 T_{\mathrm{cs}}$.
The $E''$ triplet in the inset of Fig.~\ref{fig:DOS1}a and the $A_{1}'$
triplet in the inset of Fig.~\ref{fig:DOS1}c saturate
to the same value at high field.
To avoid confusion we stress, however that in general $T_v \neq T_{\infty}$.
The most general statement is that both the temperature at which the critical field diverges and the magnitude of the dominant triplet OP at high field are determined by $\max\{T_{\mathrm{ct}},T_{\mathrm{ctz}}\}$.
In this form the above statement applies equally to both $K$- and the $\Gamma$-model considered in Sec.~\ref{sec:G_model}.

The DOS and the variation of the OPs for the two constant $T$ cuts of the phase diagram in Fig.~\ref{fig:phase_d1}b is shown in Fig.~\ref{fig:DOS1}c and Fig.~\ref{fig:DOS1}d.
The OPs along the low temperature cut crossing the first order transition line are discontinuous as shown in the inset of Fig.~\ref{fig:DOS1}c. 
Similarly, the DOS does not evolve smoothly across the first order transition. 
Nevertheless, it is not easily seen in Fig.~\ref{fig:DOS1}c.
This is explained by the smearing of the DOS as well as the smallness of the jump in the quasi-particle spectral gap across the transition. 
At higher fields the OPs as well as the shape of the DOS deep in the $s'$ phase reach saturation as can be seen from the last two DOS curves in Fig.~\ref{fig:DOS1}c corresponding to the largest fields shown.

Naturally, the DOS for the high temperature cut Fig.~\ref{fig:DOS1}d is qualitatively similar to Fig.~\ref{fig:DOS1}b.

\subsection{The \texorpdfstring{$\Gamma$}{}-model of SOC}
\label{sec:G_model}
\subsubsection{\texorpdfstring{$T_{\mathrm{ct}} > T_{\mathrm{ctz}}$}{}: dominant \texorpdfstring{$E''$}{} triplet channel}

A representative phase diagram of the Ising superconductor with nodal SOC is shown in the Fig.~\ref{fig:phase_d2}a.
Here as in the $K$-model in the same regime, $T_{\mathrm{ct}} > T_{\mathrm{ctz}}$ the OP has an $s+i f$ symmetry all over the superconducting phase provided the disorder scattering rate is not substantially smaller than $T_\mathrm{cs}$.
In clean system there is a small island of $s'$ phase separated from the $s+i f$ phase by the intermediate $s+i f +is'$ phase and at higher $T$ bordering the domain of the normal state, see Fig.~\ref{fig:phase_d3}.
Even when the disorder eliminates the $s'$ phase the critical temperature remains non-monotonic function of the field, Fig.~\ref{fig:phase_d2}a.

The DOS shown for $T< T_{\mathrm{ctz}}$ and $T> T_{\mathrm{ct}}$ in the Fig.~\ref{fig:DOS2}a and Fig.~\ref{fig:DOS2}b is characterized by the peak following the field dependent OPs and strongly broadened at elevated magnetic fields.
In addition, for fields $B > B^*$, where $B^* = \eta_0(T,B^*)$ the spectral gap is closed along the $\Gamma M$  line, where 12 nodes are formed causing topological edge states in stripe shaped samples \cite{He2018}. 
So that above the characteristic field, $B^*$ slightly exceeding the Pauli critical field obtained for $\Delta_{\mathrm{SO}}=0$, $N(\omega)$ is finite for all positive $\omega>0$.
In the clean limit it results in a $V$-shaped DOS.
At finite disorder concentration the spectral gap closes, and the $\Gamma$-model of Ising superconductor provides us with yet another example of gapless superconductivity at $B>B^*$. 
At weak disorder, $N(0) \sim \exp(- \pi \eta_0/\Gamma)$ is finite, and yet suppressed exponentially \cite{Gorkov1985}.

In general, the self-consistent Born approximation fails at low energies \cite{Graf1996}, and the detailed analysis of the DOS in this regime as well as the zero-energy Majorana surface states is beyond the scope of the present work.
Still, we note that the overall low-energy behavior of the bulk DOS stays nearly unaffected by moderately weak disorder, consistent with Fig.~\ref{fig:DOS2}.

\subsubsection{\texorpdfstring{$T_{\mathrm{ctz}} > T_{\mathrm{ct}}$}{}: dominant \texorpdfstring{$A_{1}'$}{} triplet channel}

In distinction with the $K$-model the high field first order transition between the $s+if$ and $s'$ phases proceeds via an intermediate $s+ i f + i s'$ phase, see Fig.~\ref{fig:phase_d2}b.
In the intermediate phase the OP contains two types of triplets. 
The $E''$ ($i f$) triplets are $\mathcal{T}$ odd as they are induced by the externally applied field breaking the $\mathcal{T}$ symmetry explicitly.
In contrast, $A_{1}'$ ($i s'$) triplets are present due to the spontaneous $\mathcal{T}$ symmetry breaking.
Despite this difference the triplet components have the same phase and the OP is unitary all over the phase diagram.
Although the superconducting state breaks $\mathcal{T}$ symmetry, both extrinsically and spontaneously, because the OP stays unitary there is no net Cooper pair spin polarization.

For the $\Gamma$-model with $T_{\mathrm{ctz}}<T_{\mathrm{ct}}$ the DOS curves for fixed $T=0.4T_{\mathrm{cs}}$ ($T=0.8T_{\mathrm{cs}}$) and ascending sequence of $B$ is shown in in Fig.~\ref{fig:DOS2}a (Fig.~\ref{fig:DOS2}b).
Similarly, the results for the case $T_{\mathrm{ct}}<T_{\mathrm{ctz}}$ for fixed $T=0.4T_{\mathrm{cs}}$ ($T=0.8T_{\mathrm{cs}}$) and ascending sequence of $B$ is shown in in Fig.~\ref{fig:DOS2}c (Fig.~\ref{fig:DOS2}d).

\section{Gorkov equation and qualitative considerations}
\label{sec:GE}

Our goal is to compute the DOS at any given part of the phase diagram in the $(T,B)$ plane.
This goal is achieved in two steps.
First the OPs are found by the process of minimization of the mean field thermodynamic potential.
Then with these OPs as an input we calculate the DOS by solving the Gorkov equation.
The first step is detailed in Sec.~\ref{sec:order_parameters} and covered partially in our previous work \cite{Mockli2020}.
This section is devoted to the actual calculation of the DOS.

To this end we introduce the normal and anomalous Matsubara Green functions as $2\times 2$ matrices in the spin space,
\begin{align}
G_{ss'}\left(\mathbf{r},\mathbf{r};\tau,\tau'\right) & =-\left\langle T_{\tau} \psi_{\mathbf{rs}}\left(\tau\right)\psi_{\mathbf{r}'s'}^{\dagger}\left(\tau'\right)\right\rangle, 
\notag \\
F_{ss'}\left(\mathbf{r},\mathbf{r};\tau,\tau'\right) & =-\left\langle T_{\tau}\psi_{\mathbf{rs}}\left(\tau\right)\psi_{\mathbf{r}'s'}\left(\tau'\right)\right\rangle. \label{eq:F.4.2}
\end{align}
Here, $\psi_{\mathbf{rs}}\left(\tau\right)=e^{H\tau}\psi_{\mathbf{r}s}e^{-H\tau}$
are the field operators in the Heisenberg representation, where $\tau$
is the imaginary time. 
The symbol $T_{\tau}$ stands for the time-ordering operator,
and $\left\langle \cdots\right\rangle$ indicates thermal averaging.

The disorder averaged Green function is a $4 \times 4$ matrix 
\begin{equation}
\hat{G}\left(\mathbf{k};\omega_{n}\right)=\left[\begin{array}{cc}
G\left(\mathbf{k};\omega_{n}\right) & F\left(\mathbf{k};\omega_{n}\right)\\
-F^{*}\left(-\mathbf{k};\omega_{n}\right) & -G^{*}\left(-\mathbf{k};\omega_{n}\right)
\end{array}\right], 
\label{eq:4}
\end{equation}
where we have introduced the Fourier transformed Green function
\begin{equation}
G\left(\mathbf{k};\omega_{n}\right)=\int_{V}d\mathbf{r}\int_{0}^{\beta}d\tau e^{-i\mathbf{k}\cdot\mathbf{r}+i\omega_{n}\tau}G\left(\mathbf{r};\tau\right) 
\label{eq:F.4.5}
\end{equation}
with the Matsubara frequencies $\omega_{n}=\left(2n+1\right)\pi T$ and similarly for $F\left(\mathbf{k};\omega_{n}\right)$.
The matrix $\hat{G}$ is defined in the direct product of particle-particle or Nambu and spin spaces.
This implies that each of the Green functions in Eq.~\eqref{eq:4} is a matrix in spin space in accordance with Eq.~\eqref{eq:F.4.2}.

The spin unresolved DOS is then expressed in terms of the normal Green function,
\begin{align}
N(\omega) =-\frac{N_{0}}{\pi}\sum_{s} 
\mathrm{Im} \int \mathrm{d}\xi_{k}
\left\langle G_{ss}\left(\mathbf{k};\omega+i 0^+\right)\right\rangle_{\mathrm{F}}\, ,
\label{eq:10}
\end{align}
where $0^+$ is a positive infinitesimal.

The Green function, $\hat{G}$ satisfies the Gorkov equation which can be written as 
\begin{equation}
\left[i\omega_n \hat{\sigma}_0 - \hat{H}_{\mathrm{BdG}}(\mathbf{k})-\hat{\Sigma}\left(\omega_{n}\right) \right]\hat{G}\left(\mathbf{k};\omega_n \right)=\hat{\sigma}_0\, ,
\label{eq:GE}
\end{equation}
where $\hat{\sigma}_0 = \mathrm{diag}\left(\sigma_{0},\sigma_{0}\right)$ is a unit matrix of rank 4.
The  Bogoliubov-de Gennes Hamiltonian, 
\begin{align}
\label{eq:BdG}
    \hat{H}_{\mathrm{BdG}}(\mathbf{k})\! = \!\begin{bmatrix}
\xi_{\mathbf{k}}\!+\!\left[\boldsymbol{\gamma}\!\left(\mathbf{k}\right)\!-\!\mathbf{B}\right]\!\!\cdot\!\boldsymbol{\sigma}  &   \Delta\!(\mathbf{k}) \\
\Delta^{\dagger}\!(\mathbf{k}) & 
-\xi_{\mathbf{k}}\!+\!\left[\boldsymbol{\gamma}\!\left(\mathbf{k}\right)\!+\!\mathbf{B}\right]\!\!\cdot\!\boldsymbol{\sigma}
\end{bmatrix}\!,
\end{align}
where the superconducting OP, $\Delta(\mathbf{k})$ is given by Eq.~\eqref{eq:OP}.

The effect of the disorder scattering is described within the self-consistent Born approximation by the 
self-energy appearing in Gorkov equation \eqref{eq:GE},
\begin{equation}
\hat{\Sigma}\left(\omega_{n}\right)=\Gamma\int\frac{\mathrm{d}\varphi_{\mathbf{k}}}{2\pi}
\int\frac{\mathrm{d}\xi_{\mathbf{k}}}{\pi}\hat{\sigma}_{z}
\hat{G}\left(\mathbf{k};\omega_{n}\right)\hat{\sigma}_{z},
\label{eq:Sigma}
\end{equation}
where $\Gamma = (2 \tau)^{-1}$, and $\hat{\sigma}_{z}=\mathrm{diag}\left(\sigma_{0},-\sigma_{0}\right)$.

To obtain the DOS we solved Eqs.~\eqref{eq:GE} and \eqref{eq:Sigma} numerically by the method of iterations.
With the initial guess of the self-energy the matrix inversion in Eq.~\eqref{eq:GE} gives the Green function which in turn is used in order to find an updated self-energy from Eq.~\eqref{eq:Sigma}.
These steps are repeated until convergence is reached.

Before we proceed to the discussion of the calculated DOS for various representative parts of the phase diagram we present qualitative picture of the effect of the combined action of SOC, magnetic field and triplet correlation on the DOS in the next section. 

\subsection{Qualitative picture of the DOS broadening}
\label{sec:qualit}
To understand how triplet correlations affect $N(\omega)$, 
it is useful to consider the singlet Ising superconductor first with 
the OP $\Delta(\mathbf{k}) = \eta_0 i \sigma_2$.
In this case, when $\mathbf{B}=0$, SOC does not show up due to Anderson theorem \cite{Anderson1959}.
On the other hand, for $\Delta_{\mathrm{SO}}=0$ an in-plane $B$ splits the BCS peak, without causing its broadening \cite{Maki1964}. 
Nevertheless, when both magnetic field and SOC are present the disorder causes a finite broadening of the DOS.

To clarify this we consider the unitary transformation of $H_{\mathrm{BdG}}$, Eq.~\eqref{eq:BdG} diagonalizing the normal components of $\hat{G}$.
We consider the $K$-model of SOC for simplicity.
Let us refer to the momenta such that $\hat{\gamma}(\mathbf{k})>0$,  ($\hat{\gamma}(\mathbf{k})<0$) as belonging to $+K$ and $-K$ pockets, respectively.
The diagonalization is carried out separately for momenta in $\pm K$ pockets by the unitary transformation
\begin{align}
\label{hat_U}
\hat{U} = \begin{bmatrix}
U_{\pm} & 0 \\
0 & U_{\mp}
\end{bmatrix}\, , \quad U_{\pm} = \cos\frac{\theta}{2} \mp i \sigma_2 \sin \frac{\theta}{2}\, , 
\end{align}
where the angle $\theta$ satisfies $\sin \theta = |B|/\sqrt{B^2 + \Delta_{\mathrm{SO}}^2}$. 

The BdG Hamiltonian, Eq.~\eqref{eq:BdG} is diagonalized, by the transformation
$\bar{H}_{\mathrm{BdG}} = \hat{U}   H_{\mathrm{BdG}}  \hat{U}^{-1}$, for 
$\mathbf{k}\in \pm K$,
\begin{align}
\label{eq:Hbar}
\bar{H}_{\mathrm{BdG}}\!\! =\!\!  
\begin{bmatrix}
\xi_{\mathbf{k}} \! \pm \! \sigma_3 \sqrt{B^2\! +\! \Delta_{\mathrm{SO}}^2} \!\! & \Delta'(\mathbf{k}) \\
\Delta^{\dagger '}(\mathbf{k})\!\! & \! -\! \xi_{\mathbf{k}}\!  \pm \! \sigma_3 \sqrt{B^2 \!+\! \Delta_{\mathrm{SO}}^2} 
\end{bmatrix}\! , 
\end{align}
where similar to Eq.~\eqref{eq:OP} we have 
$\Delta'(\mathbf{k})=[\psi'(\mathbf{k})\sigma_{0}+\mathbf{d}'(\mathbf{k})\cdot\boldsymbol{\sigma}]i\sigma_{2}.$
The singlet part of the transformed OP is $\eta_0' = \eta_0 \cos \theta$, and the triplet part $\mathbf{d}'(\mathbf{k}) = \mp i \eta_0\sin \theta \hat{y}$ for $\mathbf{k} \in \pm K$.

The disorder potential, Eq.~\eqref{eq:Hdis} is similarly transformed $\bar{H}_{dis} =\hat{U}   H_{dis}  \hat{U}^{-1}$,
\begin{align}
\label{eq:Hdis_Tr}
\bar{H}_{dis}& =  \sum_j \sum_{s} \sum_{\mathbf{k},\mathbf{k}'}{\vphantom{\sum}}'
u_{\mathbf{k}-\mathbf{k}'} e^{i \mathbf{R}_j(\mathbf{k}-\mathbf{k}')}    c_{\mathbf{k}s}^{\dagger}c_{\mathbf{k}'s} 
\notag \\
  & +
\sum_j \sum_{s,s'} \sum_{\mathbf{k},\mathbf{k}'}{\vphantom{\sum}}''
\left( \cos \theta \mp i \sigma_{2;ss'} \sin \theta \right) 
\notag \\
&\phantom{CC} \times u_{\mathbf{k}-\mathbf{k}'} e^{i \mathbf{R}_j(\mathbf{k}-\mathbf{k}')} \!   c_{\mathbf{k}s}^{\dagger}c_{\mathbf{k}'s'} \, ,
\end{align}
where $\sum'$ denotes the summation over the momenta $\mathbf{k}$ and $\mathbf{k}'$ belonging to the same pocket. 
This terms accounts therefore for the intra-pocket scattering.
The summation in the second term, $\sum''$ accounts for the inter-pocket scattering.
The upper (lower) sign describes the scattering from $\pm K$ to $\mp K$ pockets, respectively.

\begin{figure}
\includegraphics[scale=0.8]{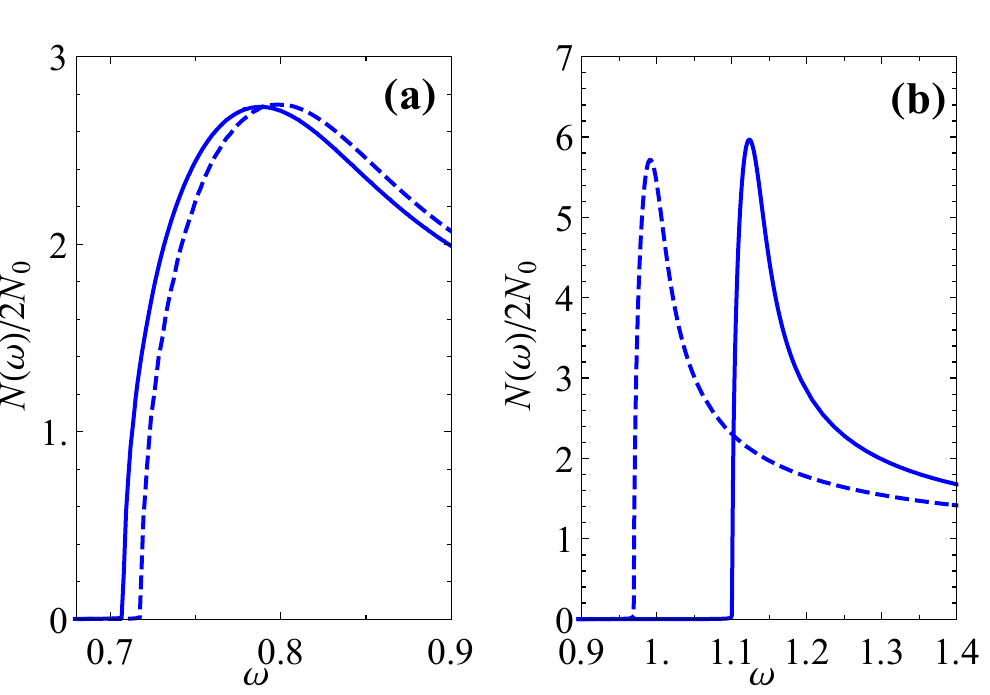}
\caption{\label{pic12} Normalized DOS, $N\left(\omega\right)/2N_{0}$ as a function of the energy $\omega$ in units of $T_{\mathrm{cs}}$ for the $K$-model. (a): solid line: DOS for original model with $B=2.5T_{\mathrm{cs}}$, $ \Delta_{\mathrm{SO}}=15T_{\mathrm{cs}}$, $\Gamma=0.5T_{\mathrm{cs}}$ and  $\left|\eta_{0}\right|=0.8T_{\mathrm{cs}}$. 
Dashed line: DOS for the effective magnetic impurity model with the effective scattering rate \eqref{eq:Gamma_eff}, and with the rest of the parameters as in the original model. 
Panel (b): Dashed line: DOS for singlet OP 
Solid line: DOS for mixed singlet-triplet OP. 
The parameters are $B=2T_{\mathrm{cs}}, \Delta_{\mathrm{SO}}=15T_{\mathrm{cs}}, \Gamma=0.1T_{\mathrm{cs}}$ and for illustration we take $\left|\eta_{0}\right|=T_{\mathrm{cs}}$ for the dashed curve and $\left|\eta_{0}\right|=T_{\mathrm{cs}},\left|\eta_{E2}\right|=T_{\mathrm{cs}}$ for the solid.}
\end{figure}

The shape of the $N(\omega)$ close to the peak, $\omega \approx \eta_0$ is determined by the states in the energy interval of the order $\eta_0$ around $E_F$.
For such low energies the bands split by the SOC as appears in Eq.~\eqref{eq:Hbar} can be considered as decoupled.
Furthermore, the inter-pocket scattering term in the transformed disorder potential, \eqref{eq:Hdis_Tr} acquires a spin flipping component, $\propto B/\Delta_{\mathrm{SO}}$. 
It appears that to the leading order in the parameter, $B/\Delta_{\mathrm{SO}} \ll 1$ the energy dependence 
of the DOS can be captures by a simple model of the anisotropic magnetic impurity scattering \cite{Mockli2020a}.
The Gorkov equation for such a problem reads
\begin{align}
    \left[i\omega_n \hat{\sigma}_0 - \hat{H}^{\mathrm{eff}}_{\mathrm{BdG}}(\mathbf{k})-\hat{\Sigma}^{\mathrm{eff}}\left(\omega_{n}\right) \right]\hat{G}\left(\mathbf{k};\omega_n \right)=\hat{\sigma}_0,
\label{eq:GE_eff}
\end{align}
where the effective BdG Hamiltonian takes a simple form, 
\begin{align}
\label{eq:BdG_eff}
\hat{H}^{\mathrm{eff}}_{\mathrm{BdG}}(\mathbf{k}) = \begin{bmatrix}
\xi_{\mathbf{k}} \sigma_0    &   \eta_0 i \sigma_2 \\
-\eta^*_0 i \sigma_2 & - \xi_{\mathbf{k}}\sigma_0
\end{bmatrix} ,
\end{align}
and the effective magnetic disorder gives rise to the self-energy,
\begin{align}
    \hat{\Sigma}^{\mathrm{eff}}\left(\omega_{n}\right)=
    \Gamma_m^{\mathrm{eff}}
    \int\frac{\mathrm{d}\varphi_{\mathbf{k}}}{2\pi}
    \int \frac{\mathrm{d}\xi_{\mathbf{k}}}{\pi}\hat{\sigma}_{2}\hat{G}\left(\mathbf{k};\omega_{n}\right)\hat{\sigma}_{2}\,  ,
    \label{eq:Sigma_eff}
\end{align}
where $\hat{\sigma}_{2}=\mathrm{diag}\left(\sigma_{2},\sigma_{2}\right)$ and the scattering off the magnetic impurities is characterized by the effective rate,
\begin{align}
\label{eq:Gamma_eff}
    \Gamma_m^{\mathrm{eff}} = \frac{1}{2} \Gamma \frac{B^2}{B^2 + \Delta_{\mathrm{SO}}^2}.
\end{align}
Even though we have considered the limit $B/\Delta_{\mathrm{SO}} \ll 1$ we retained the magnetic field in the denominator of \eqref{eq:Gamma_eff} to stress that the effective rate cannot exceed the scalar impurity scattering rate.
The prefactor of a $1/2$ is needed as only half of all the scattering events acquire a spin flipping component.
We have checked the validity of the effective model captured by Eqs.~\eqref{eq:BdG_eff} and \eqref{eq:Gamma_eff} numerically.
We have shown that the DOS given by the effective and the original models agree well, see Fig. \ref{pic12}a.

Although the effective model of magnetic impurities captures well the shape of the DOS
it does not reproduce $B_c$.
The physical reason for this is the limitation of the above effective model to energies, $\omega_n \ll \Delta_{\mathrm{SO}}$.
Indeed, when only one of the spin split bands is populated, the $B_c$ is determined by the pair breaking equation known from the theory of magnetic scattering with renormalized $T_{\mathrm{cs}}$ \cite{Sosenko2017}.
In contrast, when both spin split bands are occupied the pair breaking equation differs from that of the  to magnetic impurities model, because the frequencies $\omega_n > \Delta_{\mathrm{SO}}$ contribute to $B_c$.
For such frequencies the spin-independent Hamiltonian, \eqref{eq:BdG_eff} is inadequate since the two spin-split bands cannot be considered separately.

We now turn to the discussion of the influence of the triplet correlations on the shape of the DOS.
For definiteness we focus on the OP, \eqref{eq:OP} with finite singlet component $\eta_0$ and a finite triplet component specified by the vector $\mathbf{d} = d_2(\mathbf{k}) \hat{y}$, where $d_2(\mathbf{k})$ is an odd function of momentum.
The unitary transformation, Eq.~\eqref{hat_U} transforms the OP such that the singlet components acquires a small correction, $\eta_0' = \eta_0 \cos \theta  - i d_2 \sin \theta$.
This correction does not contribute to the broadening to the leading order. 

The triplet components encoded by $\mathbf{d}' \approx \mathbf{d}$ describe the pairs of electrons with the parallel spins.
Such states differ in energy by an amount $\Delta_{\mathrm{SO}}$.
Therefore, these terms represent a small perturbation in the Gorkov equation in the limit, $|\mathbf{d}| \ll \Delta_{\mathrm{SO}}$.
It makes it clear that except for the overall shift of the spectral peak the equal spin triplet components make a negligible contribution to its shape.
We tested this statement numerically.
As is evident from Fig. \ref{pic12}b the triplet component has little direct effect on the shape of the DOS.
Rather it changes the phase diagram as determined by the self-consistency condition, and in this way has a strong indirect effect on DOS.
More specifically, the above arguments show that as the triplet correlations make the critical field higher the broadening towards the critical field make the DOS curves progressively more broadened at elevated magnetic fields, see Fig.~\ref{fig:res1}.

\section{Thermodynamic potential and the phase diagram} 
\label{sec:order_parameters}

We write the difference of the thermodynamic potential in superconducting and normal state in the form of a Landau expansion,
\begin{align}
\label{eq:ThermP}
(V^2 N_0)^{-1} \Omega (\eta_0,\eta_A,\eta_{E1},\eta_{E2})  & = 
\Omega^{(2)} + \Omega^{(4)} \, .
\end{align}
We first present the second order terms \cite{Mockli2020}, 
\begin{align}\label{Omega2}
\Omega^{(2)}  = & C_{s_A} |\eta_0|^2 +  C_{t_A} |\eta_{A}|^2 + 
\sum_{j=1,2}C_{t_{Ej}} |\eta_{Ej}|^2 
\notag \\
& - 
2 C_{s_A,t_E} \mathrm{Im}\left\{\eta_0^*(B_x \eta_{E2} - B_y \eta_{E1})\right\}\, ,
\end{align}
where the last term describes the coupling between the $A_{1}'$ singlet and the components of the $E''$ triplet OP \cite{Mockli2019}.
The structure of this term is fixed by symmetries.
In our notations the pair of components $(\eta_{E1},\eta_{E2})$ transforms exactly as $(-B_y,B_x)$ under $D_{3h}$, and furthermore under the time reversal operation the flipping of the magnetic field is compensated by the OPs conjugation thanks to the imaginary part in Eq.~\eqref{Omega2}.

In what follows for definiteness we set $B_y =0$.
The full set of mean field equations yields $\eta_{E1}=0$ in this case.
For this reason we list only the coefficients which are not multiplying the $\eta_{E1}$ to simplify the presentation.

For the sake of clarity we present the Landau expansion of the thermodynamic potential for the $K$-model,  i.e. for the nodeless SOC. 
Employing the notations, $\omega_n = \pi T (2 n+ 1)$ for Matsubara frequency labeled by an integer $n$, and $\tilde{\omega}_n = \omega_n + \text{sgn}\left(\omega_{n}\right)\Gamma$ we have \cite{Mockli2020},
\begin{subequations}
\label{eq:Coeff_Omega2}
\begin{align}
C_{s_A}\! =\!\sum_{\omega_n >0}
\frac{ 2\pi T B^{2} \tilde{\omega}_n}
{\omega_n[\tilde{\omega}_n(B^2 +\omega_n^2)\! +\! \omega_n \Delta_{\mathrm{SO}}^2 ] }
\!+\!\ln \! \frac{T}{T_\mathrm{cs}},  
\end{align}
\begin{equation}
C_{t_{E2}}\!=\!\sum_{\omega_{n}>0}\!\!
\frac{2\pi T\left[\Gamma\left(B^{2}\!+\!\omega_{n}^{2}\right)\!+\!\omega_{n}\Delta_{\mathrm{SO}}^{2}\right]}{\omega_{n}\left[\tilde{\omega}_{n}\!\left(B^{2}\!+\!\omega_{n}^{2}\right)\!+\!\omega_{n}\Delta_{\mathrm{SO}}^{2}\right]}
+\!
\mathrm{ln}\frac{T}{T_{\mathrm{ct}}}\!, \end{equation}
\begin{equation}
C_{t_A}=\sum_{\omega_{n}>0}\frac{2\pi T\Gamma }{\omega_{n}\tilde{\omega}_{n}}
+\mathrm{ln}\frac{T}{T_{\mathrm{ctz}}},   
\end{equation}
\begin{equation}
C_{s_A,t_E}= \sum_{\omega_{n}>0}\frac{2 \pi T \Delta_{\mathrm{SO}} }{\tilde{\omega}_{n}\left(B^{2}+\omega_{n}^{2}\right)+\omega_{n}\Delta_{\mathrm{SO}}^{2}}. 
\end{equation}
\end{subequations}
%%%%%%%%%%%%%

The terms of the fourth order in the OPs can be represented in the form,
\begin{align}\label{4th_a}
\Omega^{(4)} = \Omega^{(4)}_{A} + \Omega^{(4)}_{E} + \Omega^{(4)}_{AE}\, ,
\end{align}
where the first term describes the coupling of the singlet and triplet OPs of $A_1'$ symmetry.
The second term describes the contribution of $E''$ triplets.
And finally the last term describes the coupling of OPs of different symmetry allowed at the fourth order.
In the explicit form each term of the equation \eqref{4th_a} reads
\begin{subequations}\label{4th}
\begin{align}\label{4th_b}
\Omega^{(4)}_{A}  = & 
D_1 |\eta_{0}|^{4} + D_2 |\eta_{A}|^{4}
+ D_3 |\eta_{0}|^2 |\eta_{A}|^2 
\notag \\
&
+ (D_4 \eta_{0}^{*2}\eta_{A}^{2}+c.c.),
\end{align}
\begin{align}\label{4th_c}
\Omega^{(4)}_{E}  = &  K  |\eta_{E2}|^{4} ,
\end{align}
\begin{align}\label{4th_d}
& \Omega^{(4)}_{AE} \! =\! 
L_{0E}|\eta_{0}|^2 |\eta_{E2}|^2\!  + \!L_{AE}|\eta_{E2}|^2|\eta_A|^2 
\!+\! \Big\{ L'_{0E}\eta_{0}^{*2}\eta_{E2}^{2}
\notag \\
& + L'_{AE} \eta_{E2}^{*2}\eta_{A}^{2} + M_{0E} |\eta_{0}|^2\eta_{0}^{*}\eta_{E2} 
  +  M'_{0E}\eta_{0}^{*}\eta_{E2}^{2}\eta_{E2}^{*}
  \notag \\
&  + N_1\eta_{0}^{*}\eta_{E2}^{*}\eta_{A}^{2}
  + N_2\eta_{0}^{*}\eta_{E2}\left|\eta_{A}\right|^{2}+c.c. \Big\}\, ,
\end{align}
\end{subequations}
where $c.c.$ refers to each term in the curly brackets.
All of the coefficients appearing in Eq.~\eqref{4th} are derived in the next three sections, Secs.~\ref{sec:Omega_qc_GF}, \ref{sec:Landau_exp} and \ref{sec:Eilenberger},
and presented in details in the Appendix \ref{sec:Free energy coefficients} for the $K$-model of SOC.

Below we describe the derivation of the expansion coefficients of the thermodynamic potential.
The self-consistency equations, \eqref{eq:self} are equivalent to the minimization of the thermodynamic potential, 
\begin{equation}
\Omega=-\beta^{-1}\ln\left(\mathrm{Tr}e^{-\beta\left(H_{0}+H_{\mathrm{dis}} + H_{\mathrm{MF}}\right)}\right) -\bar{H}_i - \Omega_0
\label{eq:Omega}
\end{equation}
considered as a function of the OPs \cite{Bruus2004}.
In Eq.~\eqref{eq:Omega} we have subtracted the thermodynamic potential in the normal state for convenience. 
The second term in Eq.~\eqref{eq:Omega} is readily obtained from the mean field Eq.~\eqref{eq:MF} and the interaction, Eq.~\eqref{Hint}, 
\begin{align}
\label{eq:barH_i}
\bar{H}_i & = - V^{2}N_{0}\biggr[
\mathcal{L}\left(T_{\mathrm{cs}}\right)\left|\eta_{0}\right|^{2}\\
& + \mathcal{L}\left(T_{\mathrm{ct}}\right)\underset{j=1,2}{\sum}\left|\eta_{Ej}\right|^{2}
+\mathcal{L}\left(T_{\mathrm{ctz}}\right)\left|\eta_{A}\right|^{2}\biggr]\, ,\nonumber
\end{align}
where we have introduced the notation, $\mathcal{L}(T)=\ln( 2\Lambda e^{\gamma_{E}}/T\pi)$.

To deal with the disorder we employ a quasi-classical approach \cite{Kita2015}.
To implement this approach, in the next section we express the disorder averaged thermodynamic potential, Eq.~\eqref{eq:Omega} via the quasi-classical Green function. 

\subsection{Thermodynamic potential and quasi-classical Green function}
\label{sec:Omega_qc_GF}
In the presence of the disorder it is convenient to use an alternative representation of the thermodynamic potential, Eq.~\eqref{eq:Omega}, $\Omega =\Omega' -\bar{H}_i $, where
\begin{align}
\label{eq:Int_lambda}
\Omega' = \int_{0}^{1} d \lambda\left\langle H_{\mathrm{MF}}\right\rangle_{\lambda}  \, ,
\end{align}
and $\left\langle\cdots \right\rangle _{\lambda}$ stands for the thermal averaging with respect to the quadratic Hamiltonian, $H_{0}+H_{\mathrm{dis}}+\lambda H_{\mathrm{MF}}$ with $H_{\mathrm{MF}}$ specified by Eq.~\eqref{eq:MF}.
Equivalently, at a given $\lambda$ the averaging in Eq. \eqref{eq:Int_lambda} is performed with respect to the original mean field Hamiltonian with the OP $\Delta(\mathbf{k})$ replaced by $\lambda \Delta(\mathbf{k})$.
Our derivation leading to Eq.~\eqref{eq:Omega_p3} is a variant of the original one in Ref.~\cite{Thuneberg1984} adopted to the case of multiple OPs.
The present approach in its current form has been used recently to study the effect of disorder on the phase diagram of non-superconducting systems with multiple magnetic OPs \cite{Dzero2020}.

We define the $\lambda$-dependent anomalous Green function, $F_{\lambda,ss'}\left(\mathbf{r},\mathbf{r}';\tau,\tau'\right)$
by Eq.~\eqref{eq:F.4.2} with the thermal average, 
$\left\langle\cdots\right\rangle$ replaced by $\left\langle\cdots\right\rangle _{\lambda}$. 
With these definitions Eq.~\eqref{eq:Int_lambda} takes the form,
\begin{align}
\Omega'\!=\!\frac{V}{\beta}\underset{\mathbf{k},s_{i}}{\sum}\underset{\omega_{n}}{\sum}\!\!
\int_{0}^{1}\!\!\!d\lambda\mathrm{Re}\left[\Delta_{s_{1}s_{2}}^{*}\left(\mathbf{k}\right)F_{\lambda,s_{1}s_{2}}\left(\mathbf{k};\omega_{n}\right)\right],
\label{eq:Omega_p}
\end{align}
where we have used the explicit form of the mean field Hamiltonian, Eq.~\eqref{eq:MF} and the relation,
\begin{equation}
\left\langle c_{-\mathbf{k}s_{2}}c_{\mathbf{k}s_{1}}\right\rangle _{\lambda}=\beta^{-1}\underset{\omega_{n}}{\sum}F_{\lambda,s_{1}s_{2}}\left(\mathbf{k};\omega_{n}\right).
\label{eq:F.9}
\end{equation}

To deal with the disorder we employ a quasi-classical formalism.
The quasi-classical Green's functions are defined by
\begin{align}
\label{eq:g_quasi}
\hat{g}\left(\mathbf{k}_{\mathrm{F}};\omega_{n}\right) & = \int_{-\infty}^{\infty}\frac{d\xi_{\mathbf{k}}}{\pi}i\hat{\sigma}_{3}\hat{G}\left(\mathbf{k};\omega_{n}\right) 
\\ &  =\left[\begin{array}{cc}
g\left(\mathbf{k}_{\mathrm{F}};\omega_{n}\right) & -if\left(\mathbf{k}_{\mathrm{F}};\omega_{n}\right)\\
-if^{*}\left(-\mathbf{k}_{\mathrm{F}};\omega_{n}\right) & -g^{*}\left(-\mathbf{k}_{\mathrm{F}};\omega_{n}\right)
\end{array}\right],\nonumber
\end{align}
where $\hat\sigma_3=\mathrm{diag}(\sigma_0,-\sigma_0)$. 
We parametrize the quasi-classical Green's functions in terms of Pauli matrices as
\begin{equation}
g\left(\mathbf{k}_{\mathrm{F}};\omega_{n}\right)=g_{0}\left(\mathbf{k}_{\mathrm{F}};\omega_{n}\right)\sigma_{0}+\mathbf{g}\left(\mathbf{k}_{\mathrm{F}};\omega_{n}\right)\cdot\boldsymbol{\sigma},\label{eq:14}
\end{equation}
\begin{align}
f\left(\mathbf{k}_{\mathrm{F}};\omega_{n}\right)=\left[f_{0}\left(\mathbf{k}_{\mathrm{F}};\omega_{n}\right)\sigma_{0}+\mathbf{f}\left(\mathbf{k}_{\mathrm{F}};\omega_{n}\right)\cdot\boldsymbol{\sigma}\right]i\sigma_{2}.\label{eq:15}
\end{align}
The $\lambda$-dependent quasi-classical Green functions can be defined similarly to 
Eqs.~\eqref{eq:g_quasi}, \eqref{eq:14} and \eqref{eq:15} by adding the subscript $\lambda$ to all the quantities.
For instance, we have the definition, 
\begin{equation}
 \label{eq:17.2}
 f_{\lambda}\left(\mathbf{k}_{\mathrm{F}},\omega_{n}\right)=-\int_{-\infty}^{\infty}\frac{d\xi_{\mathbf{k}}}{\pi}F_{\lambda}\left(\mathbf{k},\omega_{n}\right)    \, .
\end{equation}
Similar to Eq.~\eqref{eq:15} we parametrize
\begin{align}
f_{\lambda}\left(\mathbf{k}_{\mathrm{F}};\omega_{n}\right)\!=\!
\left[f_{\lambda,0}\left(\mathbf{k}_{\mathrm{F}};\omega_{n}\right)\sigma_{0}
+\mathbf{f}_{\lambda}\left(\mathbf{k}_{\mathrm{F}};\omega_{n}\right)\!\cdot\!\boldsymbol{\sigma}\right]i\sigma_{2}.   
\end{align}
The definition, Eq.~\eqref{eq:17.2} allows to write Eq.~\eqref{eq:Omega_p} in the form,
\begin{align}
\label{eq:Omega_p1}
\frac{\Omega'}{V^{2}N_{0}}& = -\frac{\pi}{\beta}\underset{\omega_{n}}{\sum}
\int\frac{d\varphi_{\mathbf{k}}}{2\pi}\times\\
&\int_{0}^{1}d\lambda\mathrm{Re}\left[\underset{s_{1},s_{2}}{\sum}\Delta_{s_{1}s_{2}}^{*}\left(\mathbf{k}_{\mathrm{F}}\right)
f_{\lambda,s_{1}s_{2}}
\left(\mathbf{k}_{\mathrm{F}};\omega_{n}\right)\right].\nonumber
\end{align}
For the choice of the OP specified by Eqs.~\eqref{eq:OP} and \eqref{eq:d} we have
\begin{align}
\label{Delta_f}
\frac{1}{2}&\underset{s_{1},s_{2}}{\sum}  \Delta_{s_{1}s_{2}}^{*}\left(\mathbf{k}_{\mathrm{F}}\right)f_{\lambda,s_{1}s_{2}}\left(\mathbf{k}_{\mathrm{F}};\omega_{n}\right) = \eta_{0}^{*}f_{\lambda,0}\left(\mathbf{k}_{\mathrm{F}};\omega_{n}\right)\\
+ &   \hat{\gamma}\left(\mathbf{k}_{\mathrm{F}}\right)
\left[
\underset{j=1,2}{\sum}\eta_{Ej}^{*}f_{\lambda,j}
 \left(\mathbf{k}_{\mathrm{F}};\omega_{n}\right)
+ \eta_{A}^{*}f_{\lambda,3}\left(\mathbf{k}_{\mathrm{F}};\omega_{n}\right)\right].\nonumber
\end{align}
For brevity we henceforth write $f_{0}\left(\mathbf{k}_{\mathrm{F}},\omega_{n}\right)= f_{0}$,
$\mathbf{f}\left(\mathbf{k}_{\mathrm{F}},\omega_{n}\right)=\mathbf{f}$,
$f_{0}^{*}\left(-\mathbf{k}_{\mathrm{F}},\omega_{n}\right)= f_{0}^{*}$,
$\mathbf{f}^{*}\left(-\mathbf{k}_{\mathrm{F}},\omega_{n}\right)=\mathbf{f}^{*}$
and use the same notations for $g_{0},\mathbf{g},g^{*}_{0},\mathbf{g}^{*}$.
Furthermore, naturally we extend the same notation to all the $\lambda$-dependent quasi-classical Green functions to the remainder of the paper.
With these notations, and using Eq.~\eqref{Delta_f} the expression Eq.~\eqref{eq:Omega_p1} takes the form, 
\begin{align}
\label{eq:Omega_p2}
\frac{ \Omega'}{V^{2}N_{0}} & =-\frac{2\pi}{\beta}\underset{\omega_{n}}{\sum}\int_{0}^{1}d\lambda
\mathrm{Re}\biggr[\eta_{0}^{*}\left\langle f_{\lambda,0}\right\rangle _{\mathrm{F}}
+
\nonumber\\
 & +
\eta_{A}^{*}\left\langle \hat{\gamma}\left(\mathbf{k}_{\mathrm{F}}\right)f_{\lambda,3}\right\rangle _{\mathrm{F}}
+
\underset{j=1,2}{\sum}\eta_{Ej}^{*}\left\langle\hat{\gamma}\left(\mathbf{k}_{\mathrm{F}}\right)f_{\lambda,j}\right\rangle _{\mathrm{F}}
\biggr].
\end{align}
The standard regularization, $\sum_{\omega_n} |\omega_{n}|^{-1} = (\beta/\pi) \mathcal{L}(T)$ along with the representation $\Omega =\Omega' -\bar{H}_i $, allows us to combine the expression \eqref{eq:Omega_p2} with Eq.~\eqref{eq:barH_i} to obtain for the thermodynamic potential, \eqref{eq:Omega}, 
\begin{align}
\label{eq:Omega_p3}
\frac{\Omega}{V^{2}N_{0}} & =\frac{2\pi}{\beta}\underset{\omega_{n}}{\sum}\int_{0}^{1}d\lambda\mathrm{Re}
\biggr[\eta_{0}^{*}\left(\frac{\eta_{0}}{2\left|\omega_{n}\right|}-\left\langle f_{\lambda,0}\right\rangle _{\mathrm{F}}\right)\\
 + &\eta_{A}^{*}\left(\frac{\eta_{A}}{2\left|\omega_{n}\right|}-\left\langle \hat{\gamma}\left(\mathbf{k}_{\mathrm{F}}\right)f_{\lambda,3}\right\rangle _{\mathrm{F}}\right)
\nonumber\\
 + &\underset{j=1,2}{\sum}\eta_{Ej}^{*}\left(\frac{\eta_{Ej}}{2\left|\omega_{n}\right|}-\left\langle \hat{\gamma}\left(\mathbf{k}_{\mathrm{F}}\right)f_{\lambda,j}\right\rangle _{\mathrm{F}}\right)
\Biggr]
\nonumber\\
+ & \left|\eta_{0}\right|^{2}\ln \frac{T}{T_{\mathrm{cs}}}\!+ \!\!\underset{j=1,2}{\sum}\left|\eta_{Ej}\right|^{2}\ln\frac{T}{T_{\mathrm{ct}}}
\! +\!\left|\eta_{A}\right|^{2}\ln\frac{T}{T_{\mathrm{ct}z}}.\nonumber
\end{align}
The Eq.~\eqref{eq:Omega_p3} gives an expression of the thermodynamic potential in terms of the quasi-classical Green function.
It is greatly simplified when expanded to the fourth order in OPs.
This is done in the next section. 
\subsection{Landau expansion of the thermodynamic potential}
\label{sec:Landau_exp}

In this section, we perform the expansion of the thermodynamic potential in the OPs,
$\eta_0$, $\eta_A$, $\eta_{E1}$ and $\eta_{E2}$.
For our purposes the expansion up to the fourth order suffices. 
Clearly, in order to expand the Eq.~\eqref{eq:Omega_p3} to the fourth order we need to perform the expansion of the $\lambda$-dependent quasi-classical Green functions, $f_{\lambda,0}$ and  for $\mathbf{f}_{\lambda}$ up to the third order.
Using the quasi-classical methods, in the next section we will obtain the expansion of the original, $\lambda$-independent, functions in the form 
\begin{align}
\label{eq:fg_expand}
    f_{0}&= \sum_{\nu=0}^{\infty} f_{0}^{(\nu)}, \qquad \mathbf{f}=\sum_{\nu=0}^{\infty}\mathbf{f}^{(\nu)},
    \notag \\
    g_{0}&= \sum_{\nu=0}^{\infty} g_{0}^{(\nu)}, \qquad \mathbf{g}=\sum_{\nu=0}^{\infty}\mathbf{g}^{(\nu)},
\end{align}
where the superscript $\nu$ denotes the order of expansion.
In Eq.~\eqref{eq:fg_expand} and in what follows we extend our convention of omitting the arguments $(\mathbf{k},\omega_n)$ in each of the  functions $f_{0},\mathbf{f},g_{0},\mathbf{g}$ to the expansion coefficients $f_{0}^{(\nu)},\mathbf{f}^{(\nu)},g_{0}^{(\nu)},\mathbf{g}^{(\nu)}$ appearing in Eq.~\eqref{eq:fg_expand}.

Since the dependence of $f_{\lambda,0},\mathbf{f}_{\lambda}$ on $\lambda$
originates exclusively from changing $\Delta\left(\mathbf{k}\right)$ by $\lambda\Delta\left(\mathbf{k}\right)$  
the expansion of the $\lambda$-dependent Green functions is straightforwardly related to the expansion of the 
original, $\lambda$-independent Green functions Eq.~\eqref{eq:fg_expand}, 
\begin{equation}
f_{\lambda,0}=\sum_{\nu=0}^{\infty}\lambda^{\nu}f^{\left(\nu\right)},\ \ \ \mathbf{f}_{\lambda}=\sum_{\nu=0}^{\infty}\lambda^{\nu}\mathbf{f}^{\left(\nu\right)}.
\label{eq:F.19.3}\end{equation}
Substituting \eqref{eq:F.19.3} to \eqref{eq:Omega_p3} and performing the simple integration over $\lambda$ we obtain to the second order in OPs as introduced in Eq.~\eqref{eq:ThermP},
\begin{align}
\label{eq:Omega_LE_2}
& \Omega^{(2)}  =\pi T\underset{\omega_{n}}{\sum}
\mathrm{Re}
\biggr\{\eta_{0}^{*}\left[\frac{\eta_{0}}{\left|\omega_{n}\right|}-\left\langle f_{0}^{\left(1\right)}\right\rangle _{\mathrm{F}}\right]
\\
 & +\eta_{A}^{*}\left[\frac{\eta_{A}}{\left|\omega_{n}\right|}-\left\langle \hat{\gamma}\left(\mathbf{k}_{\mathrm{F}}\right)f_{3}^{\left(1\right)}\right\rangle _{\mathrm{F}}\right]
\notag \\
 & + 
\underset{j=1,2}{\sum}\eta_{Ej}^{*}\left[\frac{\eta_{Ej}}{\left|\omega_{n}\right|}-\left\langle \hat{\gamma}\left(\mathbf{k}_{\mathrm{F}}\right)f_{j}^{\left(1\right)}\right\rangle _{\mathrm{F}}
\right]
 \Biggr\}
 \notag 
 \\
 & + \left|\eta_{0}\right|^{2}\ln \frac{T}{T_{\mathrm{cs}}}
\! +\!\left|\eta_{A}\right|^{2}\ln \frac{T}{T_{\mathrm{ct}z}}
\! +\!\!\!\underset{j=1,2}{\sum}\left|\eta_{Ej}\right|^{2}\ln \frac{T}{T_{\mathrm{ct}}}. \nonumber
\end{align}
The first and third order terms are forbidden by the gauge invariance, $\Omega^{(1)}=\Omega^{(3)}=0$. 
Technically, this follows from the vanishing of  $\left\{ f_{0}^{\left(0\right)},\mathbf{f}^{\left(0\right)}\right\} $ and $\left\{ f_{0}^{\left(2\right)},\mathbf{f}^{\left(2\right)}\right\} $, respectively, as shown in Sec.~\ref{sec:Eilenberger}.) 

To the fourth order we obtain,
\begin{align}
\label{eq:Omega_LE_4}
\Omega^{(4)}  = & - \frac{\pi T_{\mathrm{cs}}}{2}
\underset{\omega'_{n}}{\sum}\mathrm{Re}\biggr[\eta_{0}^{*}\left\langle f_{0}^{\left(3\right)}\right\rangle _{\mathrm{F}}
+ \eta_{A}^{*} \left\langle \hat{\gamma}\left(\mathbf{k}_{\mathrm{F}}\right)f_{3}^{\left(3\right)}\right\rangle _{\mathrm{F}}
\notag \\&
+\underset{j=1,2}{\sum}\eta_{Ej}^{*}\left\langle \hat{\gamma}\left(\mathbf{k}_{\mathrm{F}}\right)f_{j}^{\left(3\right)}\right\rangle _{\mathrm{F}} 
\biggr].
\end{align}
Note that unlike in Eq.~\eqref{eq:Omega_LE_2}, in  Eq.~\eqref{eq:Omega_LE_4} the temperature is set to the critical temperature, $T=T_{\mathrm{cs}}$.
This is necessary for the consistency of the Landau expansion close to $T_{\mathrm{cs}}$.
In result, the Matsubara frequencies in Eq.~\eqref{eq:Omega_LE_4}, $\omega'_{n}=\pi T_{\mathrm{cs}}\left(2n+1\right)$ are also evaluated at $T=T_{\mathrm{cs}}$.

In the next section we compute
$\{f_0^{(1)},\mathbf{f}_0^{(1)}\}$ and $\{f_0^{(3)},\mathbf{f}_0^{(3)}\}$ by solving the quasi-classical Eilenberger equation.
Together with Eqs.~\eqref{eq:Omega_LE_2} and \eqref{eq:Omega_LE_4} this will allow us to complete the Landau expansion of the thermodynamic potential.

\subsection{Solution of Eilenberger equation}
\label{sec:Eilenberger}
In this section, we introduce Eilenberger equation, \cite{Eilenberger1968,Larkin1969} and solve it for the quasi-classical Green function, Eq.~\eqref{eq:g_quasi} up the third order in the superconducting OPs.
This, according to Eqs.~\eqref{eq:Omega_LE_2} and \eqref{eq:Omega_LE_4} allow us to compute the thermodynamic potential to the fourth order.
For brevity, in this section we denote the momentum argument, $\mathbf{k}_{\mathrm{F}}$ of the quasi-classical Green functions by $\mathbf{k}$, and denote the angular average $\left\langle \ldots\right\rangle _{\mathrm{F}}$ by $\left\langle \ldots\right\rangle$ not to be confused with the thermodynamic average. 
The Eilenberger equation in the presence of SOC and Zeeman field reads \cite{Mockli2020}
\begin{equation}
\label{eq:Eil}
\left[\left(i\omega_{n}-\hat{\Sigma}\left(\omega_{n}\right)-\hat{S}\left(\mathbf{k}\right)\right)\hat{\sigma}_{3},\hat{g}\left(\mathbf{k},\omega_{n}\right)\right]=0\, ,
\end{equation}
where 
\begin{equation}
\label{eq:S}
\hat{S}\left(\mathbf{k}\right)=\left[\begin{array}{cc}
\left(\boldsymbol{\gamma}\left(\mathbf{k}\right)-\mathbf{B}\right)\cdot\boldsymbol{\sigma} & \Delta(\mathbf{k})\\
\Delta^{\dagger}(\mathbf{k}) & \left(\boldsymbol{\gamma}\left(\mathbf{k}\right)+\mathbf{B}\right)\cdot\boldsymbol{\sigma}^{\mathrm{T}}
\end{array}\right].
\end{equation}
Equation \eqref{eq:Eil} together with the normalization condition
\begin{equation}
\label{eq:18}
\hat{g}^{2}\left(\mathbf{k};\omega_{n}\right)=\hat{\sigma}_{0}
\end{equation}
determines $\hat{g}\left(\mathbf{k},\omega_{n}\right)$. 
The disorder self-energy in Eq.~\eqref{eq:Eil} is defined in Eq.~\eqref{eq:Sigma} and is expressed in terms of the quasi-classical Green function, Eq.~\eqref{eq:g_quasi} as 
\begin{align}
\label{eq:19}
\hat{\Sigma}\left(\omega_{n}\right)=-i\Gamma\left\langle \hat{g}\left(\mathbf{k};\omega_{n}\right)\right\rangle \hat{\sigma}_{3}.
\end{align}

The Eilenberger equation \eqref{eq:Eil} is written in the product of the Nambu and spin spaces the same way as the Green function, Eq.~\eqref{eq:4}.
In what follows we denote the four Nambu blocks by a pair of indices $(i,j)$, where $i,j=1,2$.
For instance, the $(1,2)$-block of the Green function, Eq.~\eqref{eq:4} is an anomalous Green function,
$F(\mathbf{k},\omega_n)$.
The $\left(1,2\right)$-block of Eq.~\eqref{eq:Eil} gives four equations which can be presented as coupled  scalar and vector equations,
\begin{subequations}
\label{eq:Eil(1,2)}
\begin{align}
\label{eq:22}2\omega_{n}f_{0} & =2i\mathbf{f}\cdot\mathbf{B}\\
 & +\psi\left(\mathbf{k}\right)\left[g_{0}+g_{0}^{*}\right]+\mathbf{d}\left(\mathbf{k}\right)\cdot\left[\mathbf{g}-\mathbf{g}^{*}\right]\nonumber \\
 & +\Gamma[-\left\langle g_{0}+g_{0}^{*}\right\rangle f_{0}+\left\langle f_{0}\right\rangle \left(g_{0}+g_{0}^{*}\right)\nonumber \\
 & -\left\langle \mathbf{g}-\mathbf{g}^{*}\right\rangle \cdot\mathbf{f}+\left\langle \mathbf{f}\right\rangle \cdot\left(\mathbf{g}-\mathbf{g}^{*}\right)],\nonumber 
\end{align}
\begin{align}
\label{eq:23}2\omega_{n}\mathbf{f} & =2if_{0}\mathbf{B}+2\boldsymbol{\gamma}\left(\mathbf{k}\right)\times\mathbf{f}+\psi\left(\mathbf{k}\right)\left[\mathbf{g}-\mathbf{g}^{*}\right]\\
 & +i\left[\mathbf{g}+\mathbf{g}^{*}\right]\times\mathbf{d}\left(\mathbf{k}\right)+\left[g_{0}+g_{0}^{*}\right]\mathbf{d}\left(\mathbf{k}\right)\nonumber \\
 & +\Gamma[-\left\langle \mathbf{g}-\mathbf{g}^{*}\right\rangle f_{0}+\left\langle f_{0}\right\rangle \left(\mathbf{g}-\mathbf{g}^{*}\right)\nonumber \\
 & -\left\langle g_{0}+g_{0}^{*}\right\rangle \mathbf{f}+\left\langle \mathbf{f}\right\rangle \left(g_{0}+g_{0}^{*}\right)\nonumber \\
 & +i\mathbf{f}\times\left\langle \mathbf{g}+\mathbf{g}^{*}\right\rangle -i\left\langle \mathbf{f}\right\rangle \times\left(\mathbf{g}+\mathbf{g}^{*}\right)].\nonumber 
\end{align}
\end{subequations}

The $\left(1,1\right)$-block of the normalization condition \eqref{eq:18} is similarly presented as
\begin{subequations}
\label{eq:Norm(1,1)}
\begin{equation}
g_{0}^{2}+\mathbf{g}^{2}=1-f_{0}f_{0}^{*}+\mathbf{f}\cdot\mathbf{f}^{*},\label{eq:24}
\end{equation}
\begin{equation}
2g_{0}\mathbf{g}=i\mathbf{f}\times\mathbf{f}^{*}+f_{0}\mathbf{f}^{*}-f_{0}^{*}\mathbf{f}.\label{eq:25}
\end{equation}
\end{subequations}

In Eq.~\eqref{eq:fg_expand} we have introduced the Landau expansion for $f_{0}$, $\mathbf{f}$, $g_{0}$ and $\mathbf{g}$ functions.
It is convenient to separately introduce the expansion of the functions $f^*_{0}$, $\mathbf{f}^*$, $g^*_{0}$ and $\mathbf{g}^*$, 
\begin{align}
\label{eq:fg_star_expand}
    f^*_{0}&= \sum_{\nu=0}^{\infty} (f_{0}^*)^{(\nu)}, \qquad \mathbf{f}^*=\sum_{\nu=0}^{\infty} (\mathbf{f}^*)^{(\nu)},
    \notag \\
    g^*_{0}&= \sum_{\nu=0}^{\infty} (g_{0}^*)^{(\nu)}, \qquad \mathbf{g}^* =\sum_{\nu=0}^{\infty}(\mathbf{g}^*)^{(\nu)},
\end{align}
where for clarity we henceforth omit the arguments $(-\mathbf{k},\omega_n)$ of the expansion coefficients,
$(f_{0}^*)^{(\nu)}$, $(\mathbf{f}^*)^{(\nu)}$, $(g_{0}^*)^{(\nu)}$ and $(\mathbf{g}^*)^{(\nu)}$ in the same way as we did for $f^*_{0}$, $\mathbf{f}^*$, $g^*_{0}$ and $\mathbf{g}^*$ before.

To solve the Eilenberger equation up to third order it suffice to consider the Eqs.~\eqref{eq:Eil(1,2)} and \eqref{eq:Norm(1,1)}.
We show in Appendix \ref{sec:Eilenberder's equations appendix} that the solutions of Eqs.~\eqref{eq:Eil(1,2)} and \eqref{eq:Norm(1,1)} found here automatically satisfy the equations contained in the 
$(1,1)$-block of Eq.~\eqref{eq:Eil} as well as the equations contained in the $(1,2)$-block of Eq.~\eqref{eq:18}.
This statement is a necessary condition for the consistency of the quasi-classical method.
Furthermore, 
the $(2,1)$- and $(2,2)$-blocks of the Eilenberger equation, \eqref{eq:Eil} are equivalent to the $(1,2)$- and $(1,1)$- blocks, respectively in the sense that the former follow from the latter by the complex conjugation and replacement of $\mathbf{k}$ by $-\mathbf{k}$. 
The same block-wise equivalence holds for the normalization condition, Eq.~\eqref{eq:18}. 
%%%%%%%%%%%%%%%%%%%%%%%%%%%%%%%%%
\subsubsection{zero order solutions}

The zero order solutions that are exact in the normal state are not fixed by the Eilenberger equation, and can be determined most easily by solving the Gorkov equation \eqref{eq:GE} directly and using the definition of the quasi-classical Green function \eqref{eq:g_quasi},
\begin{equation}
f_{0}^{(0)}=0,\ \ \ \mathbf{f}^{(0)}=0,\ \ \ g_{0}^{(0)}=\mathrm{sgn}\left(\omega_{n}\right),\ \ \ \mathbf{g}^{(0)}=0.\label{eq:Eil_0}
\end{equation}

\subsubsection{solutions to the first order}
\label{sec:Eil_1}
With the equation \eqref{eq:Eil_0} as an input, Eq.~\eqref{eq:Eil(1,2)} 
expanded to the first order reads
\begin{subequations}
\label{eq:Eil_1}
\begin{align}
\label{eq:28}
\omega_{n}f_{0}^{\left(1\right)} & =i\mathbf{f}^{\left(1\right)}\cdot\mathbf{B}+\mathrm{sgn}\left(\omega_{n}\right)\psi\left(\mathbf{k}\right)\\
 & +\Gamma\mathrm{sgn}\left(\omega_{n}\right)\left[\left\langle f_{0}^{\left(1\right)}\right\rangle -f_{0}^{\left(1\right)}\right],\nonumber 
\end{align}
\begin{align}
\omega_{n}\mathbf{f}^{\left(1\right)} & =if_{0}^{\left(1\right)}\mathbf{B}+\boldsymbol{\gamma}\left(\mathbf{k}\right)\times\mathbf{f}^{\left(1\right)}+\mathrm{sgn}\left(\omega_{n}\right)\mathbf{d}\left(\mathbf{k}\right)\label{eq:29}\\
 & +\Gamma\mathrm{sgn}\left(\omega_{n}\right)\left[\left\langle \mathbf{f}^{\left(1\right)}\right\rangle-\mathbf{f}^{\left(1\right)}\right].\nonumber 
\end{align}
\end{subequations}
To the first order the normalization condition, Eq.~\eqref{eq:Norm(1,1)} yields 
\begin{equation}
\label{eq:30} 
g_{0}^{\left(1\right)}=0,\ \ \ \mathbf{g}^{\left(1\right)}=0.
\end{equation}
Equation \eqref{eq:Eil_1} has been solved in detail in Ref.~\cite{Mockli2020}.
Here we outline the procedure used to obtain the solution as the same approach is used here to obtain  solutions at higher orders.

First, one solves Eq.~\eqref{eq:Eil_1} for $f_{0}^{\left(1\right)},\mathbf{f}^{\left(1\right)}$ with 
$\left\langle f_{0}^{\left(1\right)}\right\rangle ,\left\langle \mathbf{f}^{\left(1\right)}\right\rangle $ considered as given along with the OPs appearing in Eq.~\eqref{eq:Eil_1}.
The solutions then are averaged over angles.
This results in a linear equations for $\left\langle f_{0}^{\left(1\right)}\right\rangle ,\left\langle \mathbf{f}^{\left(1\right)}\right\rangle $ which is easily solved.
Finally, the obtained values of the latter angular averages are used to express  $f_{0}^{\left(1\right)},\mathbf{f}^{\left(1\right)}$ via the OPs.

The first order solutions obtained by following this procedure read,
\begin{subequations}
\label{eq:Eil_f1}
\begin{align}
f_{0}^{\left(1\right)} & =\frac{\Delta_{\mathrm{SO}}^{2}+\tilde{\omega}_{n}^{2}}{\left|\tilde{\omega}_{n}\right|\mathcal{K}_{1}}\left(\eta_{0}+\Gamma\left\langle f_{0}^{\left(1\right)}\right\rangle \right)+\frac{iB\text{sgn}\left(\omega_{n}\right)}{\mathcal{K}_{1}}\Gamma\left\langle f_{1}^{\left(1\right)}\right\rangle\nonumber \\
  - &\frac{iB\Delta_{\mathrm{SO}}}{\left|\tilde{\omega}_{n}\right|\mathcal{K}_{1}}\eta_{E2},
\end{align}
\begin{align}
f_{1}^{\left(1\right)} & =\frac{i\text{sgn}\left(\omega_{n}\right)B}{\mathcal{K}_{1}}\left(\eta_{0}+\Gamma\left\langle f_{0}^{\left(1\right)}\right\rangle \right)+\frac{\left|\tilde{\omega}_{n}\right|}{\mathcal{K}_{1}}\Gamma\left\langle f_{1}^{\left(1\right)}\right\rangle\nonumber \\
  -&\frac{\text{sgn}\left(\omega_{n}\right)\Delta_{\mathrm{SO}}}{\mathcal{K}_{1}}\eta_{E2},
\end{align}
\begin{align}
f_{2}^{\left(1\right)} & =\frac{iB\gamma}{\left|\tilde{\omega}_{n}\right|\mathcal{K}_{1}}\left(\eta_{0}+\Gamma\left\langle f_{0}^{\left(1\right)}\right\rangle \right)\\
 +&\frac{\text{sgn}\left(\omega_{n}\right)\Delta_{\mathrm{SO}}\hat{\gamma}\left(\mathbf{k}\right)}{\mathcal{K}_{1}}\Gamma\left\langle f_{1}^{\left(1\right)}\right\rangle +\frac{\left(\tilde{\omega}_{n}^{2}+B^{2}\right)\hat{\gamma}\left(\mathbf{k}\right)}{\left|\tilde{\omega}_{n}\right|\mathcal{K}_{1}}\eta_{E2},\nonumber
\end{align}
\begin{equation}
f_{3}^{\left(1\right)}=\frac{\hat{\gamma}\left(\mathbf{k}\right)}{\left|\tilde{\omega}_{n}\right|}\eta_{A}, 
\end{equation}
\end{subequations}
where we have defined the denominator $\mathcal{K}_{1}=B^{2}+\Delta_{\mathrm{SO}}^{2}+\tilde{\omega}_{n}^{2}$ and the angular averages are
\begin{subequations}
\begin{equation}
\label{eq:28.1}\left\langle f_{0}^{\left(1\right)}\right\rangle =\frac{\left(\Delta_{\mathrm{SO}}^{2}+\left|\tilde{\omega}_{n}\right|\left|\omega_{n}\right|\right)\eta_{0}-iB\Delta_{\mathrm{SO}}\eta_{E2}}{\left|\tilde{\omega}_{n}\right|\left(B^{2}+\omega_{n}^{2}\right)+\left|\omega_{n}\right|\Delta_{\mathrm{SO}}^{2}},
\end{equation}
\begin{equation}
\left\langle f_{1}^{\left(1\right)}\right\rangle =\frac{iB\tilde{\omega}_{n}\eta_{0}-\Delta_{\mathrm{SO}}\omega_{n}\eta_{E2}}{\left|\tilde{\omega}_{n}\right|\left(B^{2}+\omega^{2}\right)+\left|\omega_{n}\right|\Delta^{2}},
\end{equation}
\end{subequations}
and $\left\langle f_{2}^{\left(1\right)}\right\rangle =\left\langle f_{3}^{\left(1\right)}\right\rangle =0$.

Using the solutions \eqref{eq:Eil_f1} we obtain for the quantities that, together with \eqref{eq:28.1},  enter the quadratic part of the thermodynamic potential, Eq.~\eqref{eq:Omega_LE_2},
\begin{subequations}
\label{eq:Eil_f1_avg}
\begin{align}
\label{eq:28.2}
\left\langle \hat{\gamma}\left(\mathbf{k}\right)f_{2}^{\left(1\right)}\right\rangle =\frac{iB\Delta_{\mathrm{SO}}\eta_{0}+\left(B^{2}+\omega_{n}^{2}\right)\eta_{E2}}{\left|\tilde{\omega}_{n}\right|\left(B^{2}+\omega_{n}^{2}\right)+\left|\omega_{n}\right|\Delta_{\mathrm{SO}}^{2}},
\end{align}
\begin{equation}
\label{eq:28.3}
\left\langle \hat{\gamma}\left(\mathbf{k}\right)f_{3}^{\left(1\right)}\right\rangle =\frac{\eta_{A}}{\left|\tilde{\omega}_{n}\right|}.
\end{equation}
\end{subequations}
Substitution of Eq.~\eqref{eq:Eil_f1_avg} in Eq.~\eqref{eq:Omega_LE_2} reproduces Eq.~\eqref{Omega2}
with coefficients Eq.~\eqref{eq:Coeff_Omega2}.
\subsubsection{solution to the second order}
\label{sec:Eil_2}
With the solutions to the zero and first order, Eqs.~\eqref{eq:Eil_0} and \eqref{eq:30}
the Eilenberger equations \eqref{eq:Eil(1,2)} to the second order take the form,
\begin{subequations}
\label{eq:Eil_2}
\begin{align}
\label{eq:32}\omega_{n}f_{0}^{\left(2\right)}  =i\mathbf{f}^{\left(2\right)}\cdot\mathbf{B}
 +\Gamma \mathrm{sgn}\left(\omega_{n}\right)
 \left[\left\langle f_{0}^{\left(2\right)}\right\rangle-f_{0}^{\left(2\right)}\right], 
\end{align}
\begin{align}
\label{eq:33}\omega_{n}\mathbf{f}^{\left(2\right)}  = &
if_{0}^{\left(2\right)}\mathbf{B}+\boldsymbol{\gamma}\left(\mathbf{k}\right)\times\mathbf{f}^{\left(2\right)}
\notag \\
& + \Gamma\mathrm{sgn}\left(\omega_{n}\right) 
 \left[\left\langle \mathbf{f}^{\left(2\right)}\right\rangle-\mathbf{f}^{\left(2\right)} \right]. 
\end{align}
\end{subequations}

Unlike Eq.~\eqref{eq:Eil_1}, Eq.~\eqref{eq:Eil_2} is homogeneous, and as a result, has only trivial solution,
\begin{equation}
\label{eq:38} 
f_{0}^{\left(2\right)}=0,\ \ \ \mathbf{f}^{\left(2\right)}=0.
\end{equation}
To the second order, the normalization condition, Eq.~\eqref{eq:Norm(1,1)} gives the second order corrections to $g_0$ and $\mathbf{g}$ directly in terms of the first order solutions
\begin{subequations}
\label{eq:Eil_2_g}
\begin{equation}
\label{eq:34}
2\mathrm{sgn}\left(\omega_{n}\right)g_{0}^{\left(2\right)}=-f_{0}^{\left(1\right)}\left(f_{0}^{*}\right)^{\left(1\right)}+\mathbf{f}^{\left(1\right)}\cdot\left(\mathbf{f}^{*}\right)^{\left(1\right)},
\end{equation}
\begin{align}
\label{eq:35}
2\mathrm{sgn}\left(\!\omega_{n}\!\right)\!\mathbf{g}^{\left(2\right)}\!=\!
i\mathbf{f}^{\left(1\right)}\!\! \times \!\! (\! \mathbf{f}^{*} )^{\left(1\right)}
\!\!+\! \! f_{0}^{\left(1\right)}\!\left(\mathbf{f}^{*}\right)^{\left(1\right)}
\!\! -\! \! ( f_{0}^{*} )^{\left(1\right)}  \mathbf{f}^{\left(1\right)}\!\!.
\end{align}
\end{subequations}
Here we do not present Eq.~\eqref{eq:Eil_2_g} in the explicit form.
The angular averages of Eq.~\eqref{eq:Eil_2_g} are given below for illustration,
\begin{subequations}
\label{eq:Eil_2_g_avg}
\begin{align}
\label{eq:38.1}\left\langle g_{0}^{\left(2\right)}\right\rangle  & =E_{1}^{\left(0\right)}|\eta_{0}|^2+E_{2}^{(0)}|\eta_{E2}|^2+E_{3}^{\left(0\right)}|\eta_{A}|^2\\
 & +\left( E_{4}^{\left(0\right)}\eta_{0}\eta_{E2}^{*}+c.c.\right),\nonumber
\end{align}
\begin{align}
\label{eq:38.2}
& \left\langle g_{1}^{\left(2\right)}\right\rangle =E_{1}^{\left(1\right)}\left|\eta_{0}\right|^{2}+E_{2}^{\left(1\right)}\left|\eta_{E2}\right|^{2}+E_{3}^{\left(1\right)}\eta_{0}\eta_{E2}^{*}\\
& -E_{3}^{\left(1\right)*}\eta_{0}^{*}\eta_{E2}+\left\{ E_{4}^{\left(1\right)}\eta_{0}\eta_{A}^{*}+E_{5}^{\left(1\right)}\eta_{E2}\eta_{A}^{*}+c.c.\right\}\nonumber \end{align}
\end{subequations}
and $\left\langle g_{2}^{\left(2\right)}\right\rangle =\left\langle g_{3}^{\left(2\right)}\right\rangle =0$. The coefficients in Eq.~\eqref{eq:Eil_2_g_avg} are listed in Appendix \ref{sec:second order coefficients}.
A useful property,
\begin{align}
\label{eq:g_02}
    g_{0}^{*\left(2\right)}=g_{0}^{\left(2\right)}
\end{align}
follows from Eq.~\eqref{eq:34}.
\subsubsection{solution to the third order}
\label{sec:Eil_3}

To expand the Eq.~\eqref{eq:Eil(1,2)} to the third order we make use of Eqs.~\eqref{eq:Eil_0}, \eqref{eq:30}, \eqref{eq:38} and the property \eqref{eq:g_02},
\begin{subequations}
\label{eq:Eil_3}
\begin{align}
\label{eq:39}
 & 2\omega_{n}f_{0}^{\left(3\right)}=2i\mathbf{f}^{\left(3\right)}\cdot\mathbf{B}+2\psi\left(\mathbf{k}\right)g_{0}^{\left(2\right)}\\
 & +\mathbf{d}\left(\mathbf{k}\right)\cdot\left(\mathbf{g}^{\left(2\right)}-\left(\mathbf{g}^{*}\right)^{\left(2\right)}\right)\nonumber\\
 & +\Gamma\biggr[2\mathrm{sgn}\left(\omega_{n}\right)\left(\left\langle f_{0}^{\left(3\right)}\right\rangle -f_{0}^{\left(3\right)}\right)+2\left\langle f_{0}^{\left(1\right)}\right\rangle g_{0}^{\left(2\right)}\nonumber\\
 & -\left\langle \mathbf{g}^{\left(2\right)}-\left(\mathbf{g}^{*}\right)^{\left(2\right)}\right\rangle \cdot\mathbf{f}^{\left(1\right)}-2\left\langle g_{0}^{\left(2\right)}\right\rangle f_{0}^{\left(1\right)}\nonumber\\
 & +\left\langle \mathbf{f}^{\left(1\right)}\right\rangle \cdot\left(\mathbf{g}^{\left(2\right)}-\left(\mathbf{g}^{*}\right)^{\left(2\right)}\right)\biggr],\nonumber
\end{align}
\begin{align}
\label{eq:40}
 & 2\omega_{n}\mathbf{f}^{\left(3\right)}=2if_{0}^{\left(3\right)}\mathbf{B}+2\boldsymbol{\gamma}\left(\mathbf{k}\right)\times\mathbf{f}^{\left(3\right)}\\
 & +\psi\left(\mathbf{k}\right)\left(\mathbf{g}^{\left(2\right)}-\left(\mathbf{g}^{*}\right)^{\left(2\right)}\right)\nonumber\\
 & +i\left(\mathbf{g}^{\left(2\right)}+\left(\mathbf{g}^{*}\right)^{\left(2\right)}\right)\times\mathbf{d}\left(\mathbf{k}\right)+2g_{0}^{\left(2\right)}\mathbf{d}\left(\mathbf{k}\right)\nonumber\\
 & +\Gamma\biggr[\left\langle f_{0}^{\left(1\right)}\right\rangle \left(\mathbf{g}^{\left(2\right)}-\left(\mathbf{g}^{*}\right)^{\left(2\right)}\right)\nonumber\\
 & -\left\langle \mathbf{g}^{\left(2\right)}-\left(\mathbf{g}^{*}\right)^{\left(2\right)}\right\rangle f_{0}^{\left(1\right)}-2\mathbf{f}^{\left(1\right)}\left\langle g_{0}^{\left(2\right)}\right\rangle \nonumber\\
 & +2\mathrm{sgn}\left(\omega_{n}\right)\left(\left\langle \mathbf{f}^{\left(3\right)}\right\rangle -\mathbf{f}^{\left(3\right)}\right)+2\left\langle \mathbf{f}^{\left(1\right)}\right\rangle g_{0}^{\left(2\right)}\nonumber\\
 & +i\mathbf{f}^{\left(1\right)}\times\left\langle \mathbf{g}^{\left(2\right)}+\left(\mathbf{g}^{*}\right)^{\left(2\right)}\right\rangle \nonumber\\
 & -i\left\langle \mathbf{f}^{\left(1\right)}\right\rangle \times\left(\mathbf{g}^{\left(2\right)}+\left(\mathbf{g}^{*}\right)^{\left(2\right)}\right)\biggr].\nonumber
\end{align}
\end{subequations}
In order to solve Eq.~\eqref{eq:Eil_3}, for $f_{0}^{\left(3\right)},\mathbf{f}^{\left(3\right)}$ we follow the same procedure as in Sec.~\ref{sec:Eil_1} for finding the first order solutions. 
With the expressions for $f_{0}^{\left(3\right)},\mathbf{f}^{\left(3\right)}$ 
we calculate the angular averages entering the quartic part of the thermodynamic potential, Eq.~\eqref{eq:Omega_LE_4},
\begin{subequations}
\label{eq:Eil_f_ang}
\begin{align}
\label{eq:41} & \eta_{0}^{*}\left\langle f_{0}^{\left(3\right)}\right\rangle =D_{1}\left(\omega_{n}\right)\left|\eta_{0}\right|^{4}+L'_{0E}\left(\omega_{n}\right)\eta_{0}^{*2}\eta_{E2}^{2}\\
 & +\frac{1}{2}M'_{0E}\left(\omega_{n}\right)\eta_{0}^{*}\eta_{E2}^{2}\eta_{E2}^{*}+\frac{1}{2}N_{2}\left(\omega_{n}\right)\eta_{0}^{*}\eta_{E2}\left|\eta_{A}\right|^{2}\nonumber\\
 & +\left\{ M_{0E}\left(\omega_{n}\right)\eta_{0}\eta_{0}^{*2}\eta_{E2}+\frac{1}{2}c.c.\right\} +\frac{1}{2}N_{1}\left(\omega_{n}\right)\eta_{0}^{*}\eta_{E2}^{*}\eta_{A}^{2}\nonumber\\
 & +\frac{1}{2}L_{0E}\left(\omega_{n}\right)\left|\eta_{0}\right|^{2}\left|\eta_{E2}\right|^{2}+\frac{1}{2}D_{3}\left(\omega_{n}\right)\left|\eta_{0}\right|^{2}\left|\eta_{A}\right|^{2}\nonumber\\&+D_{4}\left(\omega_{n}\right)\eta_{0}^{*2}\eta_{A}^{2},\nonumber
\end{align}
%%%%%
\begin{align}
 \label{eq:42}& \eta_{E2}^{*}\left\langle \hat{\gamma}\left(\mathbf{k}\right)f_{2}^{\left(3\right)}\right\rangle =\frac{1}{2}L_{0E}\left(\omega_{n}\right)\left|\eta_{0}\right|^{2}\left|\eta_{E2}\right|^{2}\\
 & +L'_{0E}\left(\omega_{n}\right)^{*}\eta_{0}^{2}\eta_{E2}^{*2}+\frac{1}{2}N_{1}\left(\omega_{n}\right)\eta_{0}^{*}\eta_{E2}^{*}\eta_{A}^{2}\nonumber\\
 & +\frac{1}{2}M_{0E}\left(\omega_{n}\right)^{*}\eta_{0}^{2}\eta_{0}^{*}\eta_{E2}^{*}+\frac{1}{2}N_{2}\left(\omega_{n}\right)^{*}\eta_{0}\eta_{E2}^{*}\left|\eta_{A}\right|^{2}\nonumber\\
 & +\left\{ M'_{0E}\left(\omega_{n}\right)^{*}\eta_{0}\eta_{E2}\eta_{E2}^{*2}+\frac{1}{2}c.c.\right\} +K\left(\omega_{n}\right)\left|\eta_{E2}\right|^{4}\nonumber\\
 & +\frac{1}{2}L_{AE}\left(\omega_{n}\right)\left|\eta_{E2}\right|^{2}\left|\eta_{A}\right|^{2}+L'_{AE}\left(\omega_{n}\right)\eta_{E2}^{*2}\eta_{A}^{2},\nonumber
\end{align}
%%%%%%%
\begin{align}
\label{eq:43}& \eta_{A}^{*}\left\langle \hat{\gamma}\left(\mathbf{k}\right)f_{3}^{\left(3\right)}\right\rangle =D_{2}\left(\omega_{n}\right)\left|\eta_{A}\right|^{4}+\frac{1}{2}D_{3}\left(\omega_{n}\right)\left|\eta_{0}\right|^{2}\left|\eta_{A}\right|^{2}\nonumber\\
 & +D_{4}\left(\omega_{n}\right)^{*}\eta_{0}^{2}\eta_{A}^{*2}+N_{1}\left(\omega_{n}\right)^{*}\eta_{0}\eta_{E2}\eta_{A}^{*2}\\
 &+\left\{ \frac{1}{2}N_{2}\left(\omega_{n}\right)\eta_{0}^{*}\eta_{E2}\left|\eta_{A}\right|^{2}+c.c\right\} \nonumber\\
 &+\frac{1}{2}L_{AE}\left(\omega_{n}\right)\left|\eta_{E2}\right|^{2}\left|\eta_{A}\right|^{2}+L'_{AE}\left(\omega_{n}\right)^{*}\eta_{E2}^{2}\eta_{A}^{*2}.\nonumber
\end{align}
\end{subequations}
%%%%%%%
The coefficients in Eq.~\eqref{eq:Eil_f_ang} are even functions of the frequency, and are given explicitly  in Appendix \ref{sec:Free energy coefficients}. 
We multiplied the averages with suitable OP, as in the thermodynamic potential, Eq. ~\eqref{eq:Omega_LE_4}.
Substitution of Eq.~\eqref{eq:Eil_f_ang} in Eq. ~\eqref{eq:Omega_LE_4} reproduces 
Eq.~\eqref{4th} with the coefficients presented in Appendix \ref{sec:Free energy coefficients}.

\section{Discussion and Conclusions} 
\label{sec:Discussion}

\begin{figure}[htp]
\includegraphics[scale=0.8]{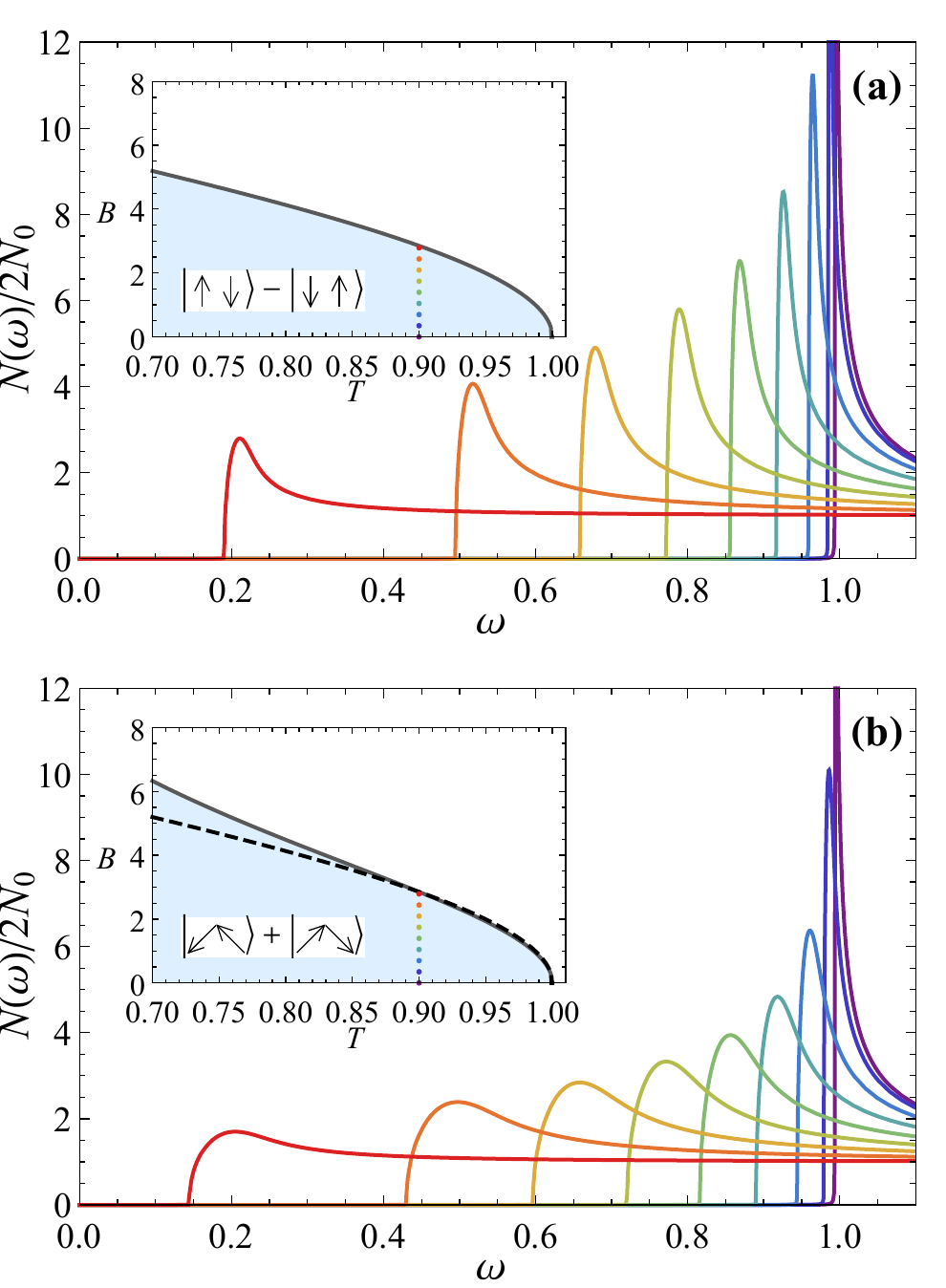}
\caption{\label{fig:conclusion}  Normalized DOS, $N\left(\omega\right)/2N_{0}$ as a function of the energy $\omega$ in units of $T_{\mathrm{cs}}$ for the $K$-model of SOC, and $\Gamma=0.1T_{\mathrm{cs}}$. 
The DOS curves broaden as the field increases.
Panel (a): purely singlet OP, $T_{\mathrm{ct}}=0$, $\Delta_{\mathrm{SO}}=15T_{\mathrm{cs}}$. 
Panel (b): mixed singlet-triplet OP,  $T_{\mathrm{ct}}=0.5 T_{\mathrm{cs}}$, $\Delta_{\mathrm{SO}}=6.28T_{\mathrm{cs}}$. 
Insets show the $(T,B)$ phase diagrams with both axes given in units of $T_{\mathrm{cs}}$.
The solid (grey) line in the insets is $B_{c}(T)$ separating the superconducting
state shown as shaded (blue) region from the normal state. 
The dashed black line in the inset of panel (b) is the is the $B_{c}(T)$ shown as the solid line in the inset to panel (a).
The coordinates of vertically aligned and evenly spaced colored dots, at $T= 0.9 T_{\mathrm{cs}} $ in insets to panel (a)[(b)] define the $(T,B)$ pairs for which the DOS is shown in the corresponding main figure using the same color scheme.  
}
\end{figure}

Triplet superconductivity remains elusive in most materials. 
Uranium-based superconductors are perhaps the only uncontroversial exception \cite{Mineev2017}. 
Some success in measuring the singlet-triplet ratio has been achieved in noncentrosymmetric superconductors by Little-Parks effect experiments \cite{Xu_PhysRevLett.124.167001}. 
Identifying triplets in DOS in noncentrosymmetric superconductors is harder and indirect, because the potentially dominant singlet component gaps out the low energy states. 
Here, we showed that excessive broadening of the coherence peaks, which cannot be explained by SOC, Zeeman effect and disorder, may be indicative of a triplet component induced by a Zeeman field. 
This behavior can be most clearly identified by observing the DOS broadening in the vicinity of the coherence peaks. 

In view of the potential importance of the unconventional paring in NbSe$_2$ based on the recent theoretical and experimental works \cite{Wickramaratne2020,Divilov2020,Hamill2020,Cho2020}  it is practically important to understand the implications of the triplet components of and OP on the tunneling data in TMDs.

In this work we have addressed the question of how the triplet correlations affect the DOS in TMD monolayers.
In practical terms, we argue that the strength of the triplet correlations is an important knob controlling the DOS evolution with an in-plane magnetic field.

We illustrate this point by a concrete example in Fig.~\ref{fig:conclusion}.
The DOS is shown for the two systems with the same critical field $B_c(T)$ and the same level of weak disorder, $\Gamma \ll T_{\mathrm{cs}}$.
In one system (Fig.~\ref{fig:conclusion}a) the order parameter is a pure singlet, while in the other (Fig.~\ref{fig:conclusion}b) system there is an admixture of the triplet component with the triplet transition temperature $T_{\mathrm{ct}} = 0.5 T_{\mathrm{cs}}$.
The critical field is the same because the SOC in the singlet superconductor, $\Delta_{\mathrm{SOC}}= 15 T_{\mathrm{cs}}$ is larger than the SOC in the superconductor of the mixed parity, $\Delta_{\mathrm{SOC}}= 6.2 T_{\mathrm{cs}}$. 
The comparison between the two figures shows that the field evolution of the DOS in the two systems is strikingly different.
The broadening is much more pronounced in the system with a small fraction of the triplet component of the OP.
The above example highlights the potential practical importance of DOS measurements in identification of triplet superconducting correlations.

In conclusion, the monolayer TMDs are promising platform for studies and controlled manipulation of triplet components of the superconducting OP.
In this work, we demonstrate that the field induced triplet correlations can be inferred from the field evolution of the tunnelling data combined with the knowledge of other parameters such as SOC and degree of system purity available from transport measurements.

\begin{acknowledgements}
We thank M. Aprili, L. Attias, T. Dvir, M. Dzero, A. Levchenko, I. Mazin, C. H. L. Quay and H. Steinberg for
useful discussions.
The authors acknowledge the financial support from the Israel Science Foundation, Grant
No. 2665/20.
D.M. acknowledges the financial support from the Swiss National Science Foundation, Project No. 184050. 
\end{acknowledgements}

\appendix

\section{Fourth order coefficients of Landau expansion of the thermodynamic potential} 
\label{sec:Free energy coefficients}
We present here the forth order coefficients of the Landau expansion of the thermodynamic potential, in the form of the summation over the Matsubara frequencies.
Here we consider the $K$-model of SOC.
As the Landau expansion is performed close to $T_{\mathrm{cs}}$ the fourth order coefficients are evaluated at $T = T_{cs}$.
Hence we present any given coefficient, $\mathcal{C}$ appearing in Eq.~\eqref{4th} in the form,
\begin{align}\label{C_omega}
\mathcal{C} = -\pi T_{\mathrm{cs}}  \sum_{\omega'_n>0}\mathcal{C}(\omega'_n)\, ,
\end{align}
where the summation is over the frequencies $\omega'_n = \pi T_{\mathrm{cs}}(2 n +1)$ labeled by an integer $n$.
We first list the coefficients in the expansion Eq.~\eqref{4th_b},
\begin{align}
D_{1}(\omega) &= \frac{\mathfrak{D}_1^{-4}}{2} 
\biggr[-\omega  \left(\omega  \tilde{\omega} +\Delta_{\mathrm{SO}}^2\right)^4  
\notag \\
+ & 2 B^2 \omega (\tilde{\omega}^2 - \Delta_{\mathrm{SO}}^2) (\Delta_{\mathrm{SO}}^2 + \
\omega \tilde{\omega} )^2
\notag \\
+ & 
B^4 ( 
3 \tilde{\omega}^4 \omega +2 \Delta_{\mathrm{SO}}^2 \tilde{\omega}^2 (\Gamma +\tilde{\omega}) - \Delta_{\mathrm{SO}}^4 \omega )
\biggr]\, ,
\end{align}
where as in the main text $\tilde{\omega} = \omega + \text{sgn}\left(\omega\right)\Gamma$, and we have defined the denominator,
\begin{align}
\mathfrak{D}_1(\omega) = B^2 \tilde{\omega} 
+\omega \left(\omega\tilde{\omega}+\Delta_{\mathrm{SO}}^2\right). 
\end{align}
The other coefficients in Eq.~\eqref{4th_b} are
\begin{equation}
D_{2}\left(\omega\right)=-\frac{\omega}{2\tilde{\omega}^{4}},
\end{equation}
\begin{align}
D_{3}\left(\omega\right)= \frac{1}{\tilde{\omega}^{2}\mathfrak{D}_{1}^{2}}
&
\biggr[B^{2}\left(\Gamma\tilde{\omega}^{2}-2\Delta_{\mathrm{SO}}^{2}\omega\right)
\notag \\ 
& -(2\omega+\Gamma)\left(\Delta_{\mathrm{SO}}^{2}+\tilde{\omega}\omega\right)^{2}
\biggr],
\end{align}
\begin{equation}
D_{4}(\!\omega \!)\!=\!\frac{-1}{2\tilde{\omega}^{2}\mathfrak{D}_{1}^{2}}
\!\!\left[B^{2}\!\!\left(\Delta_{\mathrm{SO}}^{2}\omega \!+\!\tilde{\omega}^{3}\right)
\!+\!\tilde{\omega}\left(\Delta_{\mathrm{SO}}^{2}\!+\!\tilde{\omega}\omega\right)^{2}\right]\!.
\end{equation}
The coefficient in the expansion Eq.~\eqref{4th_c} is
\begin{align}
K\left(\omega\right)&=-\frac{\omega\left(B^{2}+\omega^{2}\right)}{2\mathfrak{D}_{1}^{4}}
\big[B^{6}+B^{4}\left(2\Delta_{\mathrm{SO}}^{2}+3\omega^{2}\right)
\notag \\
& +B^{2}\left(\Delta_{\mathrm{SO}}^{4}+3\omega^{4}-4\Gamma\Delta_{\mathrm{SO}}^{2}\omega\right)
\notag \\
 &-\omega^{2}\left(3\Delta_{\mathrm{SO}}^{4}+2\Delta_{\mathrm{SO}}^{2}\left(\tilde{\omega}^{2}-\Gamma^{2}\right)-\omega^{4}\right)\big].
\end{align}
Finally, the coefficients in the expansion  Eq.~\eqref{4th_d}  are
\begin{widetext}
\begin{align}
&L_{0E}\left(\omega\right)  =\frac{1}{\mathfrak{D}_{1}^{4}}\biggr\{ B^{6}\left(\Gamma\tilde{\omega}^{2}-2\Delta_{\mathrm{SO}}^{2}\omega\right)
- B^{4}\omega\left[2\Delta_{\mathrm{SO}}^{2}\left(3\Gamma^{2}+7\Gamma\omega+6\omega^{2}\right)\!+\!4\Delta_{\mathrm{SO}}^{4}+\tilde{\omega}^{2}\omega\left(2\omega-\Gamma\right)\right]\\
  - & B^{2}\omega\left[2\Delta_{\mathrm{SO}}^{2}\omega^{2}\left(4\Gamma^{2}\!+\!10\Gamma\omega\!+\!7\omega^{2}\right)
  \!+\!\Delta^{4}\omega\left(7\Gamma+12\omega\right)\!+\! 2\Delta_{\mathrm{SO}}^{6}+\tilde{\omega}^{2}\omega^{3}\left(\Gamma+4\omega\right)\right]-\omega^{4}\left(\Gamma+2\omega\right)\left(\Delta_{\mathrm{SO}}^{2}+\tilde{\omega}\omega\right)^{2}\biggr\},\nonumber
\end{align}
\begin{equation}
L_{AE}\left(\omega\right)=\frac{-1}{\tilde{\omega}^{2}\mathfrak{D}_{1}^{2}}\left[2B^{4}\omega+B^{2}\left(\Delta_{\mathrm{SO}}^{2}(\Gamma+2\omega)+4\omega^{3}\right)-\Gamma\Delta_{\mathrm{SO}}^{2}\omega^{2}+2\omega^{5}\right],
\end{equation}
\begin{align}
L'_{0E}\left(\omega\right) & =\frac{1}{2\mathfrak{D}_{1}^{4}}\biggr\{ B^{6}\left(\Delta_{\mathrm{SO}}^{2}\omega-\tilde{\omega}^{3}\right)-B^{4}\omega\left[3\tilde{\omega}^{3}\omega-\Delta_{\mathrm{SO}}^{2}\left(\Gamma^{2}+2\Gamma\omega+2\Delta_{\mathrm{SO}}^{2}+3\omega^{2}\right)\right]\\
 & +B^{2}\omega\left[\Delta_{\mathrm{SO}}^{4}\omega(\Gamma+3\omega)-\Delta_{\mathrm{SO}}^{2}\omega^{2}\left(2\Gamma^{2}+4\Gamma\omega+\omega^{2}\right)+\Delta_{\mathrm{SO}}^{6}-3\tilde{\omega}^{3}\omega^{3}\right]-\omega^{3}\left(\Delta_{\mathrm{SO}}^{2}+\tilde{\omega}\omega\right)^{3}\biggr\},\nonumber
\end{align}
\begin{equation}
L'_{AE}(\omega)=\frac{\left(B^{2}+\omega^{2}\right)}{2\tilde{\omega}^{2}\mathfrak{D}_{1}^{2}}\left[\omega\left(B^{2}+\Delta_{\mathrm{SO}}^{2}+\omega^{2}\right)+\Gamma\Delta_{\mathrm{SO}}^{2}\right],
\end{equation}
\begin{align}
M_{0E}(\omega)=\frac{i\Delta_{\mathrm{SO}}B}{\mathfrak{D}_{1}^{4}}
&
\biggr\{ B^{4}\left[\Delta_{\mathrm{SO}}^{2}\omega-\tilde{\omega}^{2}(\Gamma+\tilde{\omega})\right]+B^{2}\omega\left[2\Delta_{\mathrm{SO}}^{4}+2\tilde{\omega}^{2}\omega^{2}+\Delta_{\mathrm{SO}}^{2}(\tilde{\omega}+\omega)^{2}\right]\\
 &+\omega\left[\omega(2\Gamma+3\omega)+\Delta_{\mathrm{SO}}^{2}\right]\left[\Delta_{\mathrm{SO}}^{2}+\tilde{\omega}\omega\right]^{2}\biggr\},\nonumber  
\end{align}
\begin{align}
M'_{0E}\left(\omega\right) & =\frac{iB\Delta_{\mathrm{SO}}\omega}{\mathfrak{D}_{1}^{4}}\biggr\{ B^{6}+B^{4}\left[2\left(\Gamma^{2}+\Delta_{\mathrm{SO}}^{2}\right)+4\Gamma\omega+5\omega^{2}\right]+B^{2}\left[\left(\Delta_{\mathrm{SO}}^{2}+\omega^{2}\right)\left(\Delta_{\mathrm{SO}}^{2}+3\omega^{2}\right)+4\tilde{\omega}^{2}\omega^{2}\right]\\
 & +\Delta_{\mathrm{SO}}^{2}\omega^{2}\left(2\omega^{2}-\Delta_{\mathrm{SO}}^{2}\right)+\omega^{4}\left(\Gamma^{2}+3\tilde{\omega}^{2}-2\Gamma\tilde{\omega}\right)\biggr\},\nonumber
\end{align}
\end{widetext}
\begin{equation}
N_{1}\left(\omega\right)=-\frac{iB\Delta_{\mathrm{SO}}}{\tilde{\omega}^{2}\mathfrak{D}_{1}^{2}}\left[\omega\left(B^{2}+\omega^{2}\right)+\Delta_{\mathrm{SO}}^{2}\tilde{\omega}\right],
\end{equation}
\begin{align}
N_{2}(\omega)=\frac{iB\Delta_{\mathrm{SO}}}{\tilde{\omega}^{2}\mathfrak{D}_{1}^{2}}
& \big[2\omega\left(B^{2}+\tilde{\omega}^{2} +\omega^{2}\right)
\notag \\
& +\Delta_{\mathrm{SO}}^{2}\left(\Gamma+2\omega\right)\big].
\end{align}

\section{Second order Landau expansion coefficients}
\label{sec:second order coefficients}

We present here the the coefficients to the averages over the second order Landau expansions from Eq.~\eqref{eq:Eil_2_g_avg}. As in the main text we denote $\tilde{\omega}_{n}=\omega_{n}+\mathrm{sgn}\left(\omega_{n}\right)\Gamma$ and define the recurring denominator
\begin{equation}
\mathcal{J}_{1}=\left|\omega_{n}\right|\Delta_{\text{SO}}^{2}+\left|\tilde{\omega}_{n}\right|\left(B^{2}+\omega_{n}^{2}\right).
\end{equation}
The coefficients are
\begin{align}
 &E_{1}^{\left(0\right)} =-\frac{\text{sgn}(\omega_{n})}{2\mathcal{J}_{1}^{2}}\biggr[2\Gamma\left|\omega_{n}\right|\left(\Delta_{\text{SO}}^{2}-B^{2}+\omega_{n}^{2}\right)\\
 & -B^{2}\left(\Gamma^{2}-\Delta_{\text{SO}}^{2}+\omega_{n}^{2}\right)+\Gamma^{2}\omega_{n}^{2}+\left(\Delta_{\text{SO}}^{2}+\omega_{n}^{2}\right){}^{2}\biggr],\nonumber
\end{align}
%-%
\begin{equation}
E_{2}^{\left(0\right)}=\frac{\text{sgn}(\omega_{n})}{2\mathcal{J}_{1}^{2}}\left[\omega_{n}^{2}\Delta_{\text{SO}}^{2}-B^{4}-B^{2}\left(\Delta_{\text{SO}}^{2}+2\omega_{n}^{2}\right)-\omega_{n}^{4}\right],
\end{equation}
\begin{equation}
E_{3}^{\left(0\right)}=-\frac{\text{sgn}(\omega_{n})}{2\left|\tilde{\omega}_{n}\right|^{2}},
\end{equation}
\begin{equation}
E_{4}^{\left(0\right)}=-\frac{i\text{sgn}\left(\omega_{n}\right)B\Delta_{\mathrm{SO}}}{2\mathcal{J}_{1}^{2}}\left(2\Gamma\left|\omega_{n}\right|+B^{2}+\Delta_{\text{SO}}^{2}+3\omega_{n}^{2}\right),
\end{equation}
\begin{equation}
E_{1}^{\left(1\right)}=-\frac{iB\left|\tilde{\omega}_{n}\right|}{\mathcal{J}_{1}^{2}}\left(\Gamma\left|\omega_{n}\right|+\Delta_{\text{SO}}^{2}+\omega_{n}^{2}\right),
\end{equation}
\begin{equation}
E_{2}^{\left(1\right)}=\frac{iB\left|\omega_{n}\right|\Delta_{\text{SO}}^{2}}{\mathcal{J}_{1}{}^{2}},
\end{equation}
\begin{equation}
E_{3}^{\left(1\right)}=\frac{\Delta_{\text{SO}}}{2\mathcal{J}_{1}^{2}}\left(\Gamma\left(B^{2}-\omega_{n}^{2}\right)-\left|\omega_{n}\right|\left(\Delta_{\text{SO}}^{2}-B^{2}+\omega_{n}^{2}\right)\right),
\end{equation}
\begin{equation}
E_{4}^{\left(1\right)}=\frac{B\text{sgn}(\omega_{n})\Delta_{\text{SO}}}{2\left|\tilde{\omega}_{n}\right|\mathcal{J}_{1}},
\end{equation}
\begin{equation}
E_{5}^{\left(1\right)}=-\frac{i\left(B^{2}+\omega_{n}^{2}\right)\text{sgn}\left(\omega_{n}\right)}{2\left|\tilde{\omega}_{n}\right|\mathcal{J}_{1}}.
\end{equation}

\section{Eilenberder equation and normalization condition: consistency of quasi-classical method}
\label{sec:Eilenberder's equations appendix}

Here we complete the discussion of Eilenberder equation, \eqref{eq:Eil}, and the normalization condition \eqref{eq:18} in Sec.~\ref{sec:Eilenberger}. 
In this appendix, as in Sec.~\ref{sec:Eilenberger}, 
we denote the momentum argument, $\mathbf{k}_{\mathrm{F}}$ of the quasi-classical Green functions by $\mathbf{k}$, and denote the angular average $\left\langle \ldots\right\rangle _{\mathrm{F}}$ by $\left\langle \ldots\right\rangle$. 
In Eq.~ \eqref{eq:Eil(1,2)} we wrote the $\left(1,2\right)$-block of Eilenberder equation, the $\left(1,1\right)$-block
is 
\begin{subequations}
\label{eq:Eil(1,1)}
\begin{align}
\label{eq:CC.1}& 0=\psi^{*}\left(\mathbf{k}\right)f_{0}-\psi\left(\mathbf{k}\right)f_{0}^{*}+\mathbf{d}\left(\mathbf{k}\right)\cdot\mathbf{f}^{*}+\mathbf{f}\cdot\mathbf{d}^{*}\left(\mathbf{k}\right)\\
& +\Gamma\left[\left\langle f_{0}^{*}\right\rangle f_{0}-\left\langle f_{0}\right\rangle f_{0}^{*}+\left\langle \mathbf{f}\right\rangle \cdot\mathbf{f}^{*}-\left\langle \mathbf{f}^{*}\right\rangle \cdot\mathbf{f}\right],\nonumber
\end{align}
\begin{align}
 \label{eq:CC.2}& 0=2\mathbf{g}\times\left(\boldsymbol{\gamma}\left(\mathbf{k}\right)-\mathbf{B}\right)-\psi^{*}\left(\mathbf{k}\right)\mathbf{f}-\mathbf{f}^{*}\psi\left(\mathbf{k}\right)\\
 &+\mathbf{d}\left(\mathbf{k}\right)f_{0}^{*} -f_{0}\mathbf{d}^{*}\left(\mathbf{k}\right)+i\mathbf{f}^{*}\times\mathbf{d}\left(\mathbf{k}\right)+i\mathbf{d}^{*}\left(\mathbf{k}\right)\times\mathbf{f}\nonumber\\
 & +\Gamma[\left\langle \mathbf{f}^{*}\right\rangle f_{0}-\left\langle f_{0}\right\rangle \mathbf{f}^{*}+\left\langle \mathbf{f}\right\rangle f_{0}^{*}-\left\langle f_{0}^{*}\right\rangle \mathbf{f}\nonumber\\
 & -i\left(\left\langle \mathbf{f}^{*}\right\rangle \times\mathbf{f}+2\mathbf{g}\times\left\langle \mathbf{g}\right\rangle +\left\langle \mathbf{f}\right\rangle \times\mathbf{f}^{*}\right)].\nonumber
\end{align}
\end{subequations}
In Eq.~\eqref{eq:Norm(1,1)} we wrote the $\left(1,1\right)$-block of the normalization
condition, the $\left(1,2\right)$-block is
\begin{subequations}
\label{eq:Norm(1,2)}
\begin{equation}
f_{0}\left(g_{0}-g_{0}^{*}\right)+\mathbf{f}\cdot\left(\mathbf{g}+\mathbf{g}^{*}\right)=0,\label{eq:CC.3}
\end{equation}
\begin{equation}
\mathbf{f}\times\left(\mathbf{g}-\mathbf{g}^{*}\right)+i\mathbf{f}\left(g_{0}-g_{0}^{*}\right)+i\left(\mathbf{g}+\mathbf{g}^{*}\right)f_{0}=0.\label{eq:CC.4}
\end{equation}
\end{subequations}
We expect the equations derived by taking the Landau expansion of Eqs.~\eqref{eq:Eil(1,1)},\eqref{eq:Norm(1,2)} up to third order to be consistent with the expressions for $f_{0}^{\left(\nu\right)},\mathbf{f}^{\left(\nu\right)},\left(f_{0}^{*}\right)^{\left(\nu\right)},\left(\mathbf{f}^{*}\right)^{\left(\nu\right)}$ and $g_{0}^{\left(\nu\right)},\mathbf{g}^{\left(\nu\right)},\left(g_{0}^{*}\right)^{\left(\nu\right)},\left(\mathbf{g}^{*}\right)^{\left(\nu\right)}$, $\nu=0,1,2$, that are found in the way illustrated in section \ref{sec:Eilenberger}. In the remainder of this appendix we perform this consistency check.  Considering the zero order terms \eqref{eq:Eil_0} and the first order terms \eqref{eq:30} the first order expansion of Eqs.~\eqref{eq:Eil(1,1)},\eqref{eq:Norm(1,2)} and the second order expansion of Eq.~\eqref{eq:Norm(1,2)} hold. The second order expansion of Eq.~\eqref{eq:Eil(1,1)} yields
\begin{subequations}
\label{eq:CC.5NN}
\begin{align}
\label{eq:CC.5}& 0=\psi^{*}\left(\mathbf{k}\right)f_{0}^{\left(1\right)}-\psi\left(\mathbf{k}\right)\left(f_{0}^{*}\right)^{\left(1\right)}+\mathbf{d}\left(\mathbf{k}\right)\cdot\left(\mathbf{f}^{*}\right)^{\left(1\right)}\\
& +\mathbf{f}^{\left(1\right)}\cdot\mathbf{d}^{*}\left(\mathbf{k}\right)+\Gamma\biggr[\left\langle \left(f_{0}^{*}\right)^{\left(1\right)}\right\rangle f_{0}^{\left(1\right)}-\left\langle f_{0}^{\left(1\right)}\right\rangle \left(f_{0}^{*}\right)^{\left(1\right)}\nonumber\\
& +\left\langle \mathbf{f}^{\left(1\right)}\right\rangle \cdot\left(\mathbf{f}^{*}\right)^{\left(1\right)}-\left\langle \left(\mathbf{f}^{*}\right)^{\left(1\right)}\right\rangle \cdot\mathbf{f}^{\left(1\right)}\biggr],\nonumber
\end{align}
\begin{align}
\label{eq:CC.6}& 0=2\mathbf{g}^{\left(2\right)}\times\left(\boldsymbol{\gamma}\left(\mathbf{k}\right)-\mathbf{B}\right)-\psi^{*}\left(\mathbf{k}\right)\mathbf{f}^{\left(1\right)}\\
& -\left(\mathbf{f}^{*}\right)^{\left(1\right)}\psi\left(\mathbf{k}\right)+\mathbf{d}\left(\mathbf{k}\right)\left(f_{0}^{*}\right)^{\left(1\right)}-f_{0}^{\left(1\right)}\mathbf{d}^{*}\left(\mathbf{k}\right)\nonumber\\
& +i\left(\mathbf{f}^{*}\right)^{\left(1\right)}\times\mathbf{d}\left(\mathbf{k}\right)+i\mathbf{d}^{*}\left(\mathbf{k}\right)\times\mathbf{f}^{\left(1\right)}+\Gamma\biggr[\left\langle \left(\mathbf{f}^{*}\right)^{\left(1\right)}\right\rangle f_{0}^{\left(1\right)}\nonumber\\
& -\left\langle f_{0}^{\left(1\right)}\right\rangle \left(\mathbf{f}^{*}\right)^{\left(1\right)}+\left\langle \mathbf{f}^{\left(1\right)}\right\rangle \left(f_{0}^{*}\right)^{\left(1\right)}-\left\langle \left(f_{0}^{*}\right)^{\left(1\right)}\right\rangle \mathbf{f}^{\left(1\right)}\nonumber\\
& -i\left(\left\langle \left(\mathbf{f}^{*}\right)^{\left(1\right)}\right\rangle \times\mathbf{f}^{\left(1\right)}+\left\langle \mathbf{f}^{\left(1\right)}\right\rangle \times\left(\mathbf{f}^{*}\right)^{\left(1\right)}\right)\biggr].\nonumber
\end{align}
\end{subequations}
The expressions for $f_{0}^{\left(1\right)},\left(f_{0}^{*}\right)^{\left(1\right)},\mathbf{f}^{\left(1\right)},\left(\mathbf{f}^{*}\right)^{\left(1\right)},\mathbf{g}^{\left(2\right)}$
derived from the procedure illustrated in section \ref{sec:Eilenberger} are consistent
with Eq.~\eqref{eq:CC.5NN}. This
can be seen by direct substitution.
Considering  Eqs.~\eqref{eq:Eil_0},\eqref{eq:30},\eqref{eq:38} the third order expansion of the normalization condition \eqref{eq:Norm(1,2)} gives
\begin{subequations}
\label{eq:CC.8NN}
\begin{equation}
\label{eq:CC.8}\mathbf{f}^{\left(1\right)}\cdot\left(\mathbf{g}^{\left(2\right)}+\left(\mathbf{g}^{*}\right)^{\left(2\right)}\right)=0,
\end{equation}
\begin{equation}
\label{eq:CC.9}
\mathbf{f}^{\left(1\right)}\!\times\!\left(\mathbf{g}^{\left(2\right)}\!-\!\left(\mathbf{g}^{*}\right)^{\left(2\right)}\right)+i\left(\mathbf{g}^{\left(2\right)}\!+\!\left(\mathbf{g}^{*}\right)^{\left(2\right)}\right)f_{0}^{\left(1\right)}=0,
\end{equation}
\end{subequations}
where we used the property~\eqref{eq:g_02}. The expressions for $f_{0}^{\left(1\right)},\mathbf{f}^{\left(1\right)},\mathbf{g}^{\left(2\right)},\left(\mathbf{g}^{*}\right)^{\left(2\right)}$
derived from the procedure illustrated in section \ref{sec:Eilenberger} are consistent
with Eq.~\eqref{eq:CC.8NN}. This
can be seen by direct substitution.
In order to complete the check with the third order expansion of Eq.~\eqref{eq:Eil(1,1)} we first take the third order expansion of Eq.~\eqref{eq:25} which together with Eqs.~\eqref{eq:Eil_0},\eqref{eq:30},\eqref{eq:38} yield $\mathbf{g}^{\left(3\right)}=0$.
Considering this result together with Eqs.~\eqref{eq:Eil_0},\eqref{eq:30},\eqref{eq:38} shows that the third order expansion of Eq.~\eqref{eq:Eil(1,1)} holds.

%\bibliographystyle{unsrt}
%\bibliography{biblio}% Produces the bibliography via BibTeX.
%\bibliographystyle{apsrev4-2}
%\bibliography{biblio}% Produces the bibliography via BibTeX.


\end{document}